\documentclass[zpreprint,zbstepj]{./LaTeX/zeus/zeus_paper}
\usepackage[english]{babel}

\newcommand{\ZcoosysB}{%
The ZEUS coordinate system is a right-handed Cartesian system, with the $Z$
axis pointing in the proton beam direction, referred to as the ``forward
direction'', and the $X$ axis pointing left towards the centre of HERA.
The coordinate origin is at the nominal interaction point.\xspace}

\newcommand{\ZcoosysfnB}{\footnote{\ZcoosysB}}

\newcommand{\Zdetdesc}{%
A detailed description of the ZEUS detector can be found 
elsewhere~\cite{zeus:1993:bluebook}. A brief outline of the 
components that are most relevant for this analysis is given
below.\xspace}
\newcommand{\Zctddesc}[1]{%
Charged particles are tracked in the central tracking detector (CTD)~\citeCTD,
which operates in a magnetic field of $1.43\Tesla$ provided by a thin 
superconducting solenoid. The CTD consists of 72~cylindrical drift chamber 
layers, organised in nine superlayers covering the polar-angle#1 region 
\mbox{$15^\circ<\theta<164^\circ$}. The transverse-momentum resolution for
full-length tracks is $\sigma(p_T)/p_T=0.0058p_T\oplus0.0065\oplus0.0014/p_T$,
with $p_T$ in $\Gev$.}
\newcommand{\Zcaldesc}{%
The high-resolution uranium--scintillator calorimeter (CAL)~\citeCAL consists 
of three parts: the forward (FCAL), the barrel (BCAL) and the rear (RCAL)
calorimeters. Each part is subdivided transversely into towers and
longitudinally into one electromagnetic section (EMC) and either one (in RCAL)
or two (in BCAL and FCAL) hadronic sections (HAC). The smallest subdivision of
the calorimeter is called a cell.  The CAL energy resolutions, as measured under
test-beam conditions, are $\sigma(E)/E=0.18/\sqrt{E}$ for electrons and
$\sigma(E)/E=0.35/\sqrt{E}$ for hadrons, with $E$ in $\Gev$.}

\chardef\usc=95
\chardef\til=126
\catcode`\@=11 
\DeclareRobustCommand\xdotspace{\futurelet\@let@token\@xdotspace}
\def\@xdotspace{%
  \ifx\@let@token.\else
  \ifx\@let@token\bgroup.\else
  \ifx\@let@token\egroup.\else
  \ifx\@let@token\/.\else
  \ifx\@let@token\ .\else
  \ifx\@let@token~.\else
  \ifx\@let@token!.\else
  \ifx\@let@token,.\else
  \ifx\@let@token:.\else
  \ifx\@let@token;.\else
  \ifx\@let@token?.\else
  \ifx\@let@token/.\else
  \ifx\@let@token'.\else
  \ifx\@let@token).\else
  \ifx\@let@token-.\else
  \ifx\@let@token\@xobeysp.\else
  \ifx\@let@token\space.\else
  \ifx\@let@token\@sptoken.\else
   .\space
   \fi\fi\fi\fi\fi\fi\fi\fi\fi\fi\fi\fi\fi\fi\fi\fi\fi\fi}
\catcode`\@=12 

\newcommand{\stru}[2]{%
   \relax\ifmmode\hbox{\vrule height#1 depth#2 width0pt}%
   \else\vrule height#1 depth#2 width0pt\fi}

\newcommand{\Ronum}[1]{\uppercase\expandafter{\romannumeral#1}}
\newcommand{\ronum}[1]{\expandafter{\romannumeral#1}}
\DeclareRobustCommand{\LaTeXZ}{%
  \LaTeX\kern-.05em4\kern-.1em
  {\raisebox{-0.2ex}{$\scriptstyle\text{ZEUS}$}}\xspace}

\DeclareMathAlphabet{\mathbf}{OT1}{cmr}{bx}{sl}
\newcommand{\eVdist}{\kern-0.06667em}

\newcommand{\Gev}{{\text{Ge}\eVdist\text{V\/}}}

\newcommand{\gev}{{\,\text{Ge}\eVdist\text{V\/}}}

\newcommand{\Tesla}{\,\text{T}}

\newcommand{\slashfrac}[2]{%
  \raisebox{0.5ex}{\ensuremath #1}\kern-0.12em/\kern-0.08em
  \raisebox{-.8ex}{\ensuremath #2}}

\newcommand{\sqr}[3]{%
    {\vcenter{\hrule height.#3ex\hbox{\vrule width.#2ex height#1ex
     \kern#1ex\vrule width.#3ex}\hrule height.#2ex}}}

\catcode`\@=11 
\newcommand{\parenbar}{\mathpalette\p@renb@r}
\def\p@renb@r#1#2{\vbox{%
  \ifx#1\scriptscriptstyle \dimen@.7em\dimen@ii.2em\else
  \ifx#1\scriptstyle \dimen@.8em\dimen@ii.25em\else
  \dimen@1em\dimen@ii.4em\fi\fi \offinterlineskip
  \ialign{\hfill##\hfill\cr
    \vbox{\hrule width\dimen@ii}\cr
    \noalign{\vskip-.3ex}%
    \hbox to\dimen@{$\mathchar300\hfil\mathchar301$}\cr
    \noalign{\vskip-.3ex}%
    $#1#2$\cr}}}
\catcode`\@=12 

\newcommand{\IP}{{\rm I$\kern-0.01667em$P}\xspace}

\mathchardef\qsm=63
\mathchardef\pls=43
\mathchardef\mns=512
\mathchardef\plm=518
\mathchardef\eql=61
\mathchardef\smallleft=300
\mathchardef\smallright=301
\mathchardef\les=316
\mathchardef\gre=318
\mathchardef\leq=532
\mathchardef\grq=533

\catcode`\@=11 
\newcounter{pict@width}
\newcounter{pict@height}
\newlength{\pict@scale}
\newcommand{\psfigadd}[4]{%
\setcounter{pict@width}{1*\ratio{#2+\pict@scale/2}{\pict@scale}}
\setcounter{pict@height}{1*\ratio{#3+\pict@scale/2}{\pict@scale}}
\setlength{\unitlength}{\pict@scale}
\hbox to #2{\hspace{-\fill}\begin{picture}(\thepict@width,\thepict@height)
\put(0,0){\psfig{figure=#1,width=#2,height=#3,clip=}}
\SetScale{0.283466457}
\SetWidth{1.763889}
{#4}
\end{picture}}
}
\newcounter{pict@widthfst}
\newcounter{pict@widthscd}
\newcounter{pict@widthtot}
\newcommand{\psfigaddtwo}[7]{%
\setcounter{pict@widthfst}{1*\ratio{#2+\pict@scale/2}{\pict@scale}}
\setcounter{pict@widthscd}{1*\ratio{#2+#4+\pict@scale/2}{\pict@scale}}
\setcounter{pict@widthtot}{1*\ratio{#2+#4+#6+\pict@scale/2}{\pict@scale}}
\setcounter{pict@height}{1*\ratio{#3+\pict@scale/2}{\pict@scale}}
\setlength{\unitlength}{\pict@scale}
\hbox{\hspace{-\fill}\begin{picture}(\thepict@widthtot,\thepict@height)
\put(0,0){\psfig{figure=#1,width=#2,height=#3,clip=}}
\put(\thepict@widthscd,0){\psfig{figure=#5,width=#6,height=#3,clip=}}
\SetScale{0.283466457}
\SetWidth{1.763889}
{#7}
\end{picture}}
}
\newcommand{\psfigror}[4]{%
\setcounter{pict@width}{1*\ratio{#2+\pict@scale/2}{\pict@scale}}
\setcounter{pict@height}{1*\ratio{#3+\pict@scale/2}{\pict@scale}}
\setlength{\unitlength}{\pict@scale}
\hbox{\begin{picture}(\thepict@width,\thepict@height)
\put(0,\thepict@height){\psfig{figure=#1,width=#3,height=#2,clip=,angle=270}}
\SetScale{0.283466457}
\SetWidth{1.763889}
{#4}
\end{picture}}
}
\newcommand{\psfigrol}[4]{%
\setcounter{pict@width}{1*\ratio{#2+\pict@scale/2}{\pict@scale}}
\setcounter{pict@height}{1*\ratio{#3+\pict@scale/2}{\pict@scale}}
\setlength{\unitlength}{\pict@scale}
\hbox{\begin{picture}(\thepict@width,\thepict@height)
\put(0,0){\psfig{figure=#1,width=#3,height=#2,clip=,angle=90}}
\SetScale{0.283466457}
\SetWidth{1.763889}
{#4}
\end{picture}}
}
\catcode`\@=12 
\newlength\listtextwidth

\catcode`\@=11 
\newlength{\@tabfninsert}
\newlength{\@tabfnwidth}
\newcommand{\tabfootnote}[2]{%
  \setlength{\@tabfninsert}{0.8em}
  \setlength{\@tabfnwidth}{\textwidth}
  \addtolength{\@tabfnwidth}{-\@tabfninsert}
  \addtolength{\@tabfnwidth}{-0.4em}
  \noindent\makebox[\@tabfninsert][r]{\footnotesize$^{#1}$\hfil}\hfill%
  \parbox[t]{\@tabfnwidth}{\footnotesize #2\hfill}}
\catcode`\@=12 

\def\citeCTD{{\cite{%
nim:a279:290,*npps:b32:181,*nim:a338:254%
}}\xspace}
\def\citeCAL{{\cite{%
nim:a309:77,*nim:a309:101,*nim:a321:356,*nim:a336:23%
}}\xspace}

\newcommand{\dspm}      {\mbox{$D^{\ast \pm}$} }
\newcommand{\dz}        {\mbox{$D^{0}$}}
\newcommand{\ptds}      {\mbox{$p_T(D^{\ast})$}}
\newcommand{\etads}     {\mbox{$\eta(D^{\ast})$}}
\newcommand{\qsq}       {\mbox{${Q^2}$}}
\newcommand{\ftwo}      {\mbox{$F_{2}$}}
\newcommand{\ftwoccb}   {\mbox{$F_{2}^{c\bar c}$}}
\begin{document}
\title{
\vspace{-5cm}
\begin{flushleft} {\normalsize \tt DESY 03-115}\\ \vspace{-.25cm}{\normalsize \tt August 2003} \end{flushleft}\vspace{2cm}
Measurement of $D^{*\pm}$  production in deep \\ inelastic $e^\pm p$
scattering at HERA
}                                                       
                    
\author{ZEUS Collaboration}
\draftversion{Post reading}
\date{}

\abstract{
Inclusive production of \dspm(2010) mesons in deep inelastic scattering has been 
measured with the ZEUS detector at HERA using an integrated luminosity of 
\mbox{81.9 pb$^{-1}$}. The decay channel \mbox{$D^{\ast +}\rightarrow D^0 \pi^+ $} 
with $D^0\to K^-\pi^+$ and corresponding antiparticle decay were used to identify 
$D^*$  mesons. Differential $D^*$ cross sections with 
\mbox{$1.5<Q^2<1000\,$GeV$^2$} and \mbox{$0.02<y<0.7$} in the kinematic region 
\mbox{$1.5<p_T(D^*)<15$\,GeV} and \mbox{$\left|\,\eta(D^*)\,\right|<1.5$} are 
compared to different QCD calculations incorporating different parameterisations 
of the parton densities in the proton. The data show sensitivity to the gluon 
distribution in the proton and are reasonably well described by 
next-to-leading-order QCD with the ZEUS NLO QCD fit used as the input parton 
density in the proton. The observed cross section is extrapolated to the full 
kinematic region in $p_T(D^*)$ and $\eta(D^*)$ in order to determine the 
open-charm contribution, \ftwoccb$(x,Q^2)$, to the proton structure function, 
$F_2$. Since, at low $Q^2$, the uncertainties of the data are comparable to those 
from the QCD fit, the measured differential cross sections in $y$ and $Q^2$ should 
be used in future fits to constrain the gluon density.

\vspace{0.5cm}

}

\makezeustitle

\def\3{\ss}                                                                                        
\pagenumbering{Roman}                                                                              
                                                   %
\begin{center}                                                                                     
{                      \Large  The ZEUS Collaboration              }                               
\end{center}                                                                                       
  S.~Chekanov,                                                                                     
  M.~Derrick,                                                                                      
  D.~Krakauer,                                                                                     
  J.H.~Loizides$^{   1}$,                                                                          
  S.~Magill,                                                                                       
  B.~Musgrave,                                                                                     
  J.~Repond,                                                                                       
  R.~Yoshida\\                                                                                     
 {\it Argonne National Laboratory, Argonne, Illinois 60439-4815}, USA~$^{n}$                       
\par \filbreak                                                                                     
  M.C.K.~Mattingly \\                                                                              
 {\it Andrews University, Berrien Springs, Michigan 49104-0380}, USA                               
\par \filbreak                                                                                     
  P.~Antonioli,                                                                                    
  G.~Bari,                                                                                         
  M.~Basile,                                                                                       
  L.~Bellagamba,                                                                                   
  D.~Boscherini,                                                                                   
  A.~Bruni,                                                                                        
  G.~Bruni,                                                                                        
  G.~Cara~Romeo,                                                                                   
  L.~Cifarelli,                                                                                    
  F.~Cindolo,                                                                                      
  A.~Contin,                                                                                       
  M.~Corradi,                                                                                      
  S.~De~Pasquale,                                                                                  
  P.~Giusti,                                                                                       
  G.~Iacobucci,                                                                                    
  A.~Margotti,                                                                                     
  A.~Montanari,                                                                                    
  R.~Nania,                                                                                        
  F.~Palmonari,                                                                                    
  A.~Pesci,                                                                                        
  G.~Sartorelli,                                                                                   
  A.~Zichichi  \\                                                                                  
  {\it University and INFN Bologna, Bologna, Italy}~$^{e}$
% ...
\newpage                                                                                           
$^{\    1}$ also affiliated with University College London, London, UK \\                          
$^{\    2}$ on leave of absence at University of                                                   
Erlangen-N\"urnberg, Germany\\                                                                     
% ...
$^{\    5}$ PPARC Advanced fellow \\                                                               
% ...
$^{\    7}$ now at Dongshin University, Naju, Korea \\                                             
% ...
$^{  15}$ on leave of absence at Columbia Univ., Nevis Labs., N.Y., US                             
A\\                                                                                                
% ...
                                                           %
                                                           %
\newpage   
                                                           %
                                                           %
\begin{tabular}[h]{rp{14cm}}                                                                       
$^{a}$ &  supported by the Natural Sciences and Engineering Research                               
          Council of Canada (NSERC) \\                                                             
$^{b}$ &  supported by the German Federal Ministry for Education and                               
          Research (BMBF), under contract numbers HZ1GUA 2, HZ1GUB 0, HZ1PDA 5, HZ1VFA 5\\         
$^{c}$ &  supported by the MINERVA Gesellschaft f\"ur Forschung GmbH, the                          
          Israel Science Foundation, the U.S.-Israel Binational Science                            
          Foundation and the Benozyio Center                                                       
          for High Energy Physics\\                                                                
$^{d}$ &  supported by the German-Israeli Foundation and the Israel Science                        
          Foundation\\                                                                             
$^{e}$ &  supported by the Italian National Institute for Nuclear Physics (INFN) \\                
$^{f}$ &  supported by the Japanese Ministry of Education, Culture,                                
          Sports, Science and Technology (MEXT) and its grants for                                 
          Scientific Research\\                                                                    
$^{g}$ &  supported by the Korean Ministry of Education and Korea Science                          
          and Engineering Foundation\\                                                             
$^{h}$ &  supported by the Netherlands Foundation for Research on Matter (FOM)\\                   
$^{i}$ &  supported by the Polish State Committee for Scientific Research,                         
          grant no. 620/E-77/SPB/DESY/P-03/DZ 117/2003-2005\\                                      
$^{j}$ &  partially supported by the German Federal Ministry for Education                         
          and Research (BMBF)\\                                                                    
$^{k}$ &  partly supported by the Russian Ministry of Industry, Science                            
          and Technology through its grant for Scientific Research on High                         
          Energy Physics\\                                                                         
$^{l}$ &  supported by the Spanish Ministry of Education and Science                               
          through funds provided by CICYT\\                                                        
$^{m}$ &  supported by the Particle Physics and Astronomy Research Council, UK\\                   
$^{n}$ &  supported by the US Department of Energy\\                                               
$^{o}$ &  supported by the US National Science Foundation\\                                        
$^{p}$ &  supported by the Polish State Committee for Scientific Research,                         
          grant no. 112/E-356/SPUB/DESY/P-03/DZ 116/2003-2005,2 P03B 13922\\                       
$^{q}$ &  supported by the Polish State Committee for Scientific Research,                         
          grant no. 115/E-343/SPUB-M/DESY/P-03/DZ 121/2001-2002, 2 P03B 07022\\                    
\end{tabular}                                                                                      
                                                           %
                                                           %

\pagenumbering{arabic} 
\pagestyle{plain}
\section{Introduction}
\label{sec-int}

Charm quarks are produced copiously in deep inelastic scattering (DIS) at HERA. 
At sufficiently high photon virtualities, $Q^2$, the production of charm 
quarks constitutes up to $30\%$ of the total cross 
section~\cite{epj:c12:35,pl:b528:199}. Previous measurements of $D^*$ cross 
sections~\cite{epj:c12:35,pl:b528:199,pl:b407:402,np:b545:21} indicate that the 
production of charm quarks in DIS in the range $1 < Q^2 < 600$ GeV$^2$ is 
consistent with calculations in Quantum Chromodynamics (QCD) in which charm is 
produced through the boson-gluon-fusion (BGF) mechanism. This implies that the 
charm cross section is directly sensitive to the gluon density in the proton.

In this paper, measurements of the $D^*$ cross section are presented with 
improved precision and in a kinematic region extending to higher $Q^2$ than the 
previous ZEUS results~\cite{epj:c12:35}. Single differential cross sections 
have been measured as a function of $Q^2$ and the Bjorken scaling variable, $x$. 
Cross sections have also been measured in two $Q^2$ ranges as a function of 
transverse momentum, $p_T(D^*)$, and pseudorapidity, $\eta(D^*)$, of the $D^*$ 
meson. The cross sections are compared to the predictions of leading-logarithmic 
Monte Carlo (MC) simulations and to a next-to-leading-order (NLO) QCD calculation 
using various parton density functions (PDFs) in the proton. In particular, the 
data are compared to calculations using the recent ZEUS NLO QCD 
fit~\cite{pr:d67:012007}, in which the parton densities in the proton are 
parameterised by performing fits to inclusive DIS measurements from ZEUS and 
fixed-target experiments. The cross-section measurements are used to extract the 
charm contribution, $F_2^{c\bar{c}}$, to the proton structure function, $F_2$.

\section{Experimental set-up}
\label{sec-exp}

The analysis was performed with data taken from 1998 to 2000, when HERA 
collided electrons or positrons with energy $E_e =$ 27.5~GeV with protons of 
energy $E_p =$ 920~GeV. The results are based on $e^-p$ and $e^+p$ samples 
corresponding to integrated luminosities of $16.7\pm 0.3$ pb$^{-1}$ and 
$65.2\pm 1.5$ pb$^{-1}$, respectively.\footnote{Hereafter, both electrons and 
positrons are referred to as electrons, unless explicitly stated otherwise.}

\Zdetdesc

\Zctddesc\ZcoosysfnB

\Zcaldesc

Presamplers (PRES)~\cite{nim:a382:419,*magill:bpre} are mounted in front of 
FCAL, BCAL and RCAL. They consist of scintillator tiles which detect particles 
originating from showers in the material between the interaction point and the 
calorimeter. This information was used to correct the energy of the scattered 
electron. The position of electrons scattered close to the electron beam 
direction is determined by a scintillator strip detector 
(SRTD)~\cite{nim:a401:63}. The SRTD signals resolve single minimum-ionising 
particles and provide a transverse position resolution of 3~mm.

The luminosity was measured from the rate of the bremsstrahlung process 
$ep~\rightarrow~e\gamma p$, where the photon was measured in a lead--scintillator
calorimeter~\cite{desy-92-066,*zfp:c63:391,*acpp:b32:2025} placed in the HERA 
tunnel at $Z=-107~{\rm m}$.

A three-level trigger system was used to select events 
online~\cite{zeus:1993:bluebook,proc:chep:1992:222}. At the third level, events 
with both a reconstructed $D^*$ candidate and a scattered-electron candidate were 
kept for further analysis. The efficiency of the online $D^*$ reconstruction, 
determined relative to an inclusive DIS trigger, was generally above $95\%$.

\section{Theoretical predictions}
\label{sec:theory}

A variety of models to describe charm production in DIS have been constructed, 
based on many theoretical ideas. A comparison of the data with these models is 
complicated by the need to produce predictions for the limited range of 
acceptance of the detector in $p_T(D^*)$ and $\eta(D^*)$. The calculation used 
in this paper to compare with the measured cross sections is based on NLO QCD 
as described in Section~\ref{sec:nlo}. Monte Carlo models also provide 
calculations in the measured kinematic region; those used are discussed in 
Section~\ref{sec:mc}. Predictions of other models are briefly discussed in 
Section~\ref{sec:models}. Most of these models only predict the total cross 
sections and cannot therefore be directly compared with the current data.

\subsection{NLO QCD calculations}
\label{sec:nlo}

The NLO predictions for $c\bar{c}$ cross sections were obtained using the 
HVQDIS program~\cite{pr:d57:2806} based on the so-called fixed-flavour-number 
scheme (FFNS). In this scheme, only light quarks ($u, d, s$) are included in 
the initial-state proton as partons whose distributions obey the DGLAP 
equations\cite{sovjnp:15:438,*sovjnp:20:94,*np:b126:298,*jetp:46:641}, and the 
$c\bar{c}$ is produced via the BGF mechanism~\cite{np:b452:109,*pl:b353:535} 
with NLO corrections~\cite{np:b392:162,*np:b392:229}. The presence of the two 
large scales, $Q^2$ and $m_c^2$, can spoil the convergence of the perturbative 
series because the neglected terms of orders higher than $\alpha_s^2$ contain 
log$(Q^2/m_c^2)$ factors which can become large. Therefore, the results of 
HVQDIS are expected to be most accurate at $Q^2 \approx m_c^2$ and to become 
less reliable when $Q^2 \gg m_c^2$. 

The following inputs have been used to obtain the predictions for $D^*$ 
production at NLO using the program HVQDIS. The recent ZEUS NLO QCD global 
fit~\cite{pr:d67:012007} to structure-function data was used as the 
parameterisation of the proton PDFs. This fit was 
repeated~\cite{misc:www:zeus2002} in the FFNS, in which the PDF has three 
active quark flavours in the proton, and $\Lambda^{(3)}_{\rm QCD}$ is set to 
0.363~GeV. In this fit, the mass of the charm quark was set to 1.35~GeV; the 
same mass was therefore used in the HVQDIS calculation of the predictions. 
The renormalisation and factorisation scales were set to 
$\mu = \sqrt{Q^2+4m_c^2}$ for charm production both in the fit and in the 
HVQDIS calculation. The charm fragmentation to a $D^*$ is carried out using 
the Peterson function~\cite{pr:d27:105}. The hadronisation fraction, 
$f(c \to D^*)$, taken from combined $e^+e^-$ measurements, was set to 
0.235~\cite{hep-ex-9912064} and the Peterson parameter, $\epsilon$, was set 
to 0.035\cite{np:b565:245}. The production cross section for charmonium 
states at HERA is larger than in high-energy $e^+e^-$ collisions. The effect 
of $J/\psi$ production on the hadronisation fraction was estimated from 
data~\cite{epj:c25:41,epj:c6:603} to be about $2\%$ and was neglected. 

As an alternative to the Peterson fragmentation function, corrections were 
applied to the partons in the NLO calculation using the {\sc Aroma} MC 
program~\cite{cpc:101:1997:135} (see Section~\ref{sec:mc}) which uses the Lund 
string fragmentation~\cite{prep:97:31}, modified for heavy quarks according to 
Bowler~\cite{zfp:c11:169,*np:b70:93}, and leading-logarithmic parton showers. 
This correction was applied on a bin-by-bin basis to the NLO calculation for 
each cross section measured, according to the formula 
$d\sigma(D^*)_{\rm NLO+MC} = d\sigma(c\bar{c})_{\rm NLO} \cdot C_{\rm had}$ 
where $C_{\rm had} = d\sigma(D^*)_{\rm MC}/d\sigma(c\bar{c})_{\rm MC}$. The 
shapes of the differential cross sections calculated at the parton level of the 
{\sc Aroma} model agreed reasonably well with those calculated from the HVQDIS 
program. The effect of the choice of hadronisation scheme is discussed in 
Sections~\ref{sec:xsec} and~\ref{sec:extract}.

To estimate the contribution of beauty production, the NLO calculation and 
hadronisation from the MC were combined, using  
$d\sigma(b \to D^*)_{\rm NLO+MC} = d\sigma(b\bar{b})_{\rm NLO} \cdot C_{\rm had}$ 
where $C_{\rm had} = d\sigma(b \to D^*)_{\rm MC}/d\sigma(b\bar{b})_{\rm MC}$. 
The ZEUS NLO QCD fit was used as the proton PDF, so that the mass used in this fit, 
$m_b=4.3$~GeV, was also used in the HVQDIS program and $\mu$ was set to 
$\sqrt{Q^2+4m_b^2}$. The hadronisation fraction, $f(b \to D^*)$, was set to 
0.173~\cite{epj:c1:439}.

An alternate way to describe charm production in QCD is the 
variable-flavour-number scheme (VFNS)\cite{zfp:c74:463,epj:c18:547}. In these 
calculations, an attempt is made to treat the heavy quarks correctly for all 
$Q^2$. Therefore, at low $Q^2$, charm is produced dynamically through the BGF 
process as in the FFNS, whereas, at higher $Q^2$, heavy-quark parton densities 
are introduced. The transition between the two extremes is treated in different 
ways by different authors~\cite{zfp:c74:463,epj:c18:547}. The ZEUS NLO QCD fit 
has been performed in this scheme using the formalism of Roberts and 
Thorne~\cite{jp:g25:1307,epj:c19:339}. Predictions from such calculations are, 
however, only available for the total charm cross section; no calculation of 
$D^*$ production in the measured kinematic range is available. 

\subsection{Monte Carlo models of charm production}
\label{sec:mc}

The MC programs {\sc Aroma} and {\sc Cascade}~\cite{epj:c19:351} were also 
compared with the measured differential cross sections. In the {\sc Aroma} MC 
program, charm is produced via the BGF process. Higher-order QCD effects are 
simulated in the leading-logarithmic approximation with initial- and 
final-state radiation obeying DGLAP evolution. The mass of the charm quark was 
set to 1.5~GeV and the proton PDF chosen was CTEQ5F3~\cite{epj:c12:375}. The 
{\sc Cascade} MC model takes a different approach to the generation of the 
hard sub-process, in which heavy-quark production is simulated in the framework 
of the semi-hard or $k_T$-factorisation 
approach~\cite{prep:100:1,*prep:189:267,*sovjnp:53:657,*sovjnp:54:867,*pl:b242:97,*np:b360:3,*np:b386:215,np:b366:135}. The matrix element used in {\sc Cascade} is the off-shell 
LO BGF process~\cite{np:b366:135,epj:c24:425}. The {\sc Cascade} initial-state 
radiation is based on CCFM 
evolution~\cite{np:b296:49,*pl:b234:339,*np:b336:18,*np:b445:49}, which 
includes $\ln (1/x)$ terms in the perturbative expansion in addition to the 
$\ln Q^2$ terms used in DGLAP evolution. To simulate final-state radiation, 
{\sc Cascade} uses \mbox{{\sc Pythia} 5.7~\cite{cpc:82:74}}. The cross section 
is calculated by convoluting the off-shell BGF matrix element with the 
unintegrated gluon density of the proton obtained from the CCFM fit to the HERA 
$F_2$ data~\cite{np:b470:3} with $m_c = 1.5$~GeV. For both {\sc Aroma} and 
{\sc Cascade}, the Lund string model is used for the fragmentation into hadrons,  
and $f(c \to D^*)$ was set to 0.235.

\subsection{Other predictions of charm production}
\label{sec:models}

The extraction of $F_2^{c\bar{c}}$ performed in this paper (see 
Section~\ref{sec:extract}) is model dependent and comparisons of $F_2^{c\bar{c}}$ 
to the predictions of models other than that used to produce it are not in 
general valid. Thus, only the FFNS model, which was used to extract 
$F_2^{c\bar{c}}$, was compared to the data. 

Several models of charm production~\cite{pl:b437:408,*pl:b470:243,*pl:b550:160,*pr:d59:014017,*yf:63:1682,*yf:63:1682,*pr:d62:074013,*yf:66:801,*yf:60:1680} 
were compared in the $x$ and $Q^2$ range of the measurements in this paper. As 
most only predict total cross sections, the comparison was performed for 
$F_2^{c\bar{c}}$. All models show similar trends, with differences typically 
less than $20\%$. Since the differences are smaller than the current precision of 
the $D^*$ cross-section measurements, these models are not considered further.

\section{Kinematic reconstruction and event selection}
\label{sub:ds}

The kinematic variables $Q^2$, $x$ and the fraction of the electron energy 
transferred to the proton in its rest frame, $y$, can be reconstructed using a 
variety of methods, whose accuracy depends on the variable of interest and its 
range: 

\begin{itemize}

\item 
for the electron method (specified with the subscript $e$), the measured energy 
and angle of the scattered lepton are used; 

\item
the double angle (DA) method~\cite{proc:hera:1991:23,*hoeger} relies on the angles of 
the scattered lepton and the hadronic energy flow; 

\item
the Jacquet-Blondel (JB) method~\cite{proc:epfacility:1979:391} is based 
entirely on measurements of the hadronic system;    

\item
the $\Sigma$-method~\cite{nim:a361:197} uses both the scattered-lepton energy and 
measurements of the hadronic system.    

\end{itemize}

The reconstruction of $Q^2$ and $x$ was performed using the $\Sigma$-method, 
since it has better resolution at low $Q^2$ than the DA method. At high $Q^2$, 
the $\Sigma$-method and the DA method are similar, and both have better 
resolution than the electron method.  

The events were selected~\cite{epj:c21:443,epj:c12:35} by the following cuts:

\begin{itemize}

\item 
the scattered electron was identified using a neural-network 
procedure~\cite{nim:a365:508,*nim:a391:360}. Its energy, $E_{e^{'}}$, was 
required to be larger than 10~GeV;

\item
$y_{e}\> \leq\> 0.95$;

\item
$y_{\mathrm{JB}}\> \geq\>  0.02$;

\item
$40\> \leq\>  \delta \>\leq\> 60$ GeV, where $\delta=\sum E_i(1-\cos\theta_i)$ 
and   $E_i$ is the energy of the calorimeter cell $i$. The sum runs over all 
cells;  

\item
a primary vertex position determined from the tracks fitted to the vertex in 
the range $|Z_{\rm vertex}| < 50$ cm; 

\item
the impact point ($X$, $Y$) of the scattered lepton on the RCAL must lie 
outside the region \mbox{26 $\times$ 14 cm$^2$} centred on $X=Y=0$.  

\end{itemize}

The angle of the scattered  lepton was determined using either its impact 
position on the CAL inner face or a reconstructed track in the CTD. The 
SRTD information was used, when available. The energy of the scattered lepton 
was corrected using the PRES, with additional corrections for non-uniformity 
due to geometric effects caused by cell and module boundaries. The quantity 
$\delta$ was calculated from a combination of CAL clusters and tracks measured 
in the CTD. The contribution to $\delta$ from the scattered lepton was 
evaluated separately after all corrections were applied as described above.

The selected kinematic region was \mbox{$1.5 < Q^2 < 1000$ GeV$^2$} and 
\mbox{$0.02 < y < 0.7$}.

\section{Selection of $D^*$ candidates}
\label{sub:dssel}

The $D^*$ mesons were identified using the decay channel 
$D^{*+}\to D^0\pi^{+}_s$ with the subsequent decay $D^0\to K^-\pi^+$ and the 
corresponding antiparticle decay, where $\pi^+_s$ refers to a low-momentum 
(``slow'') pion accompanying the $D^0$.

Charged tracks measured by the CTD and assigned to the primary event vertex were 
selected. The transverse momentum was required to be greater than 0.12~GeV. Each 
track was required to reach at least the third superlayer of the CTD. These 
restrictions ensured that the track acceptance and momentum resolution were 
high. Tracks in the CTD with opposite charges and transverse momenta 
$p_T > 0.4\gev$ were combined in pairs to form $D^0$ candidates. The tracks were 
alternately assigned the masses of a kaon and a pion and the invariant mass of 
the pair, $M_{K\pi}$, was found. Each additional track, with charge opposite to 
that of the kaon track, was assigned the pion mass and combined with the 
$D^0$-meson candidate to form a $D^*$ candidate.  
 
The signal regions for the reconstructed masses, $M(D^0)$ and 
$\Delta M=(M_{K\pi\pi_s} - M_{K\pi})$, were \mbox{$1.80 < M(D^0)<1.92$ GeV} and 
\mbox{$0.143 < \Delta M < 0.148$ GeV}, respectively. To allow the background to 
be determined, $D^0$ candidates with wrong-sign combinations, in which both 
tracks forming the $D^0$ candidates have the same charge and the third track has 
the opposite charge, were also retained. The same kinematic restrictions were 
applied as for those $D^0$ candidates with correct-charge combinations. 

\begin{figure}
\begin{center}
\mbox{\epsfig{file=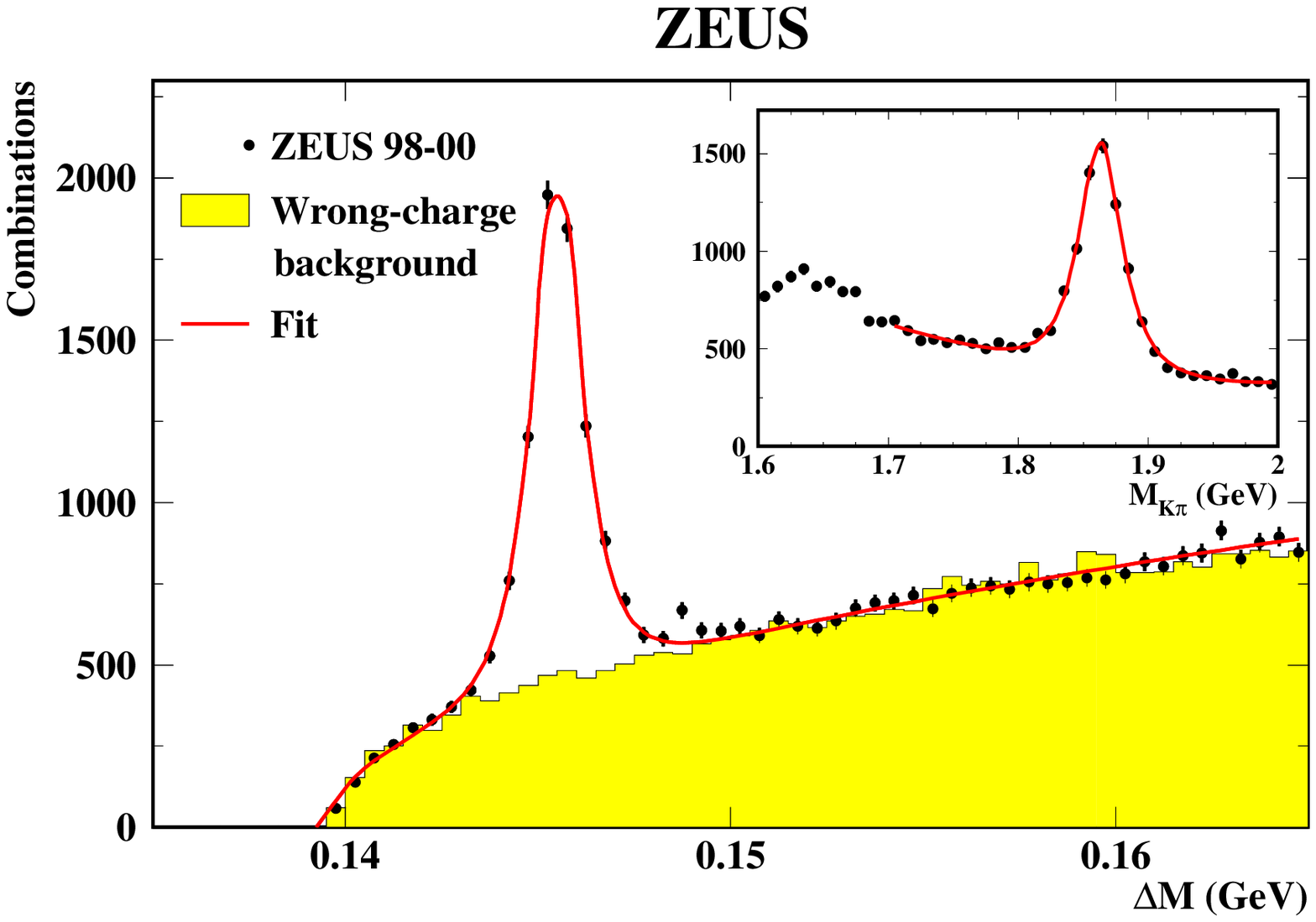,height=10.5cm}}
\caption[ig0_pap]
{The distribution of the mass difference, $\Delta M=(M_{K\pi\pi_s} - M_{K\pi})$, 
for $D^*$ candidates (solid dots). The $\Delta M$ distribution from wrong-charge 
combinations, normalised in the region $0.15 < \Delta M < 0.165$~GeV,  
is shown as the histogram. The solid line shows the result of the fit described 
in the text. The $M_{K\pi}$ distribution for the $D^0$ candidates in the range 
$0.143 < \Delta M < 0.148$~GeV is shown as an inset. The fit is the sum of a 
modified Gaussian 
to describe the signal and a second-order polynomial to describe the background.}  
\label{ig0_pap}
\end{center}
\end{figure}

The kinematic region for $D^*$ candidates was $1.5 < p_T(D^*)<15$ GeV and 
\mbox{$|\eta (D^*)|<1.5$}. Figure~\ref{ig0_pap} shows the $\Delta M$ 
distribution for the $D^*$ candidates together with the background from the 
wrong-charge combinations. The fit to the distribution has the form

$$
F =p_1 \cdot 
\exp \left( -0.5\cdot x^{1+\frac{1}{1+0.5x}} \right) + 
p_4\cdot (\Delta M - m_\pi)^{p_5} ,
$$

where $x=|(\Delta M-p_2)/p_3|$, $p_1-p_5$ are free parameters and $m_\pi$ is the 
pion mass. The ``modified'' Gaussian was used to fit the mass peak since it gave 
a better $\chi^2$ value than the conventional Gaussian form for a MC 
sample of $D^*$ mesons. 
The fit gives a peak at \mbox{$145.49 \pm 0.02$(stat.) MeV} compared with the 
PDG value of \mbox{$145.421\pm 0.010$ MeV~\cite{pr:d66:010001}}. The measured 
peak position differs from the PDG value. However, it was not corrected for detector 
effects and the systematic uncertainty was not determined. The fitted width of 
\mbox{$0.61 \pm 0.02$ MeV} is consistent with the experimental resolution. Consistent 
results were also found for the $e^+p$ and $e^-p$ data separately. For the 
range \mbox{$0.143 < \Delta M < 0.148$ GeV}, a clear signal of $D^0$ candidates 
is also shown in Figure~\ref{ig0_pap}.

The number of $D^*$ candidates determined in the two signal regions and after 
subtracting the background 
estimated from the wrong-charge sample was $5545 \pm 129$. The normalisation 
factor of the wrong-charge sample was determined as the ratio of events with 
correct-charge combinations to wrong-charge combinations in the region 
\mbox{$150<\Delta M<165$ MeV}. This factor is compatible with unity for both 
$e^-p$ and $e^+p$ data. The normalisation factors were determined for each 
bin in order to calculate the differential cross sections using the 
background-subtraction method.

\section{Acceptance corrections}
\label{sec:evsim}
  
The acceptances were calculated using the {\sc Rapgap 2.08}~\cite{cpc:86:147} 
and {\sc Herwig 6.1}~\cite{hep-ph-9912396,*cpc:67:465} MC models. The 
{\sc Rapgap} MC model was interfaced with {\sc Heracles 4.6.1}~\cite{cpc:69:155} 
in order to incorporate first-order electroweak corrections. The generated 
events were then passed through a full simulation of the detector, separately 
for $e^-p$ and $e^+p$ running, using 
{\sc Geant 3.13}~\cite{tech:cern-dd-ee-84-1} and processed and selected with 
the same programs as used for the data. 
 
The MC models were used to produce charm by the 
BGF process only. The GRV94-LO~\cite{zfp:c67:433} PDF for the proton was used, 
and the charm-quark mass was set to 1.5~GeV. The {\sc Herwig} MC contains 
leading-logarithmic parton showers whereas for {\sc Rapgap} MC, the 
colour-dipole model~\cite{cpc:71:15} as implemented in 
{\sc Ariadne 4.03}~\cite{cpc:71:15} was used  to simulate QCD 
radiation. Charm fragmentation is implemented using either the Lund string 
fragmentation ({\sc Rapgap}) or a cluster fragmentation~\cite{np:b238:492} 
model ({\sc Herwig}).

\begin{figure}
\begin{center}
\mbox{\epsfig{file=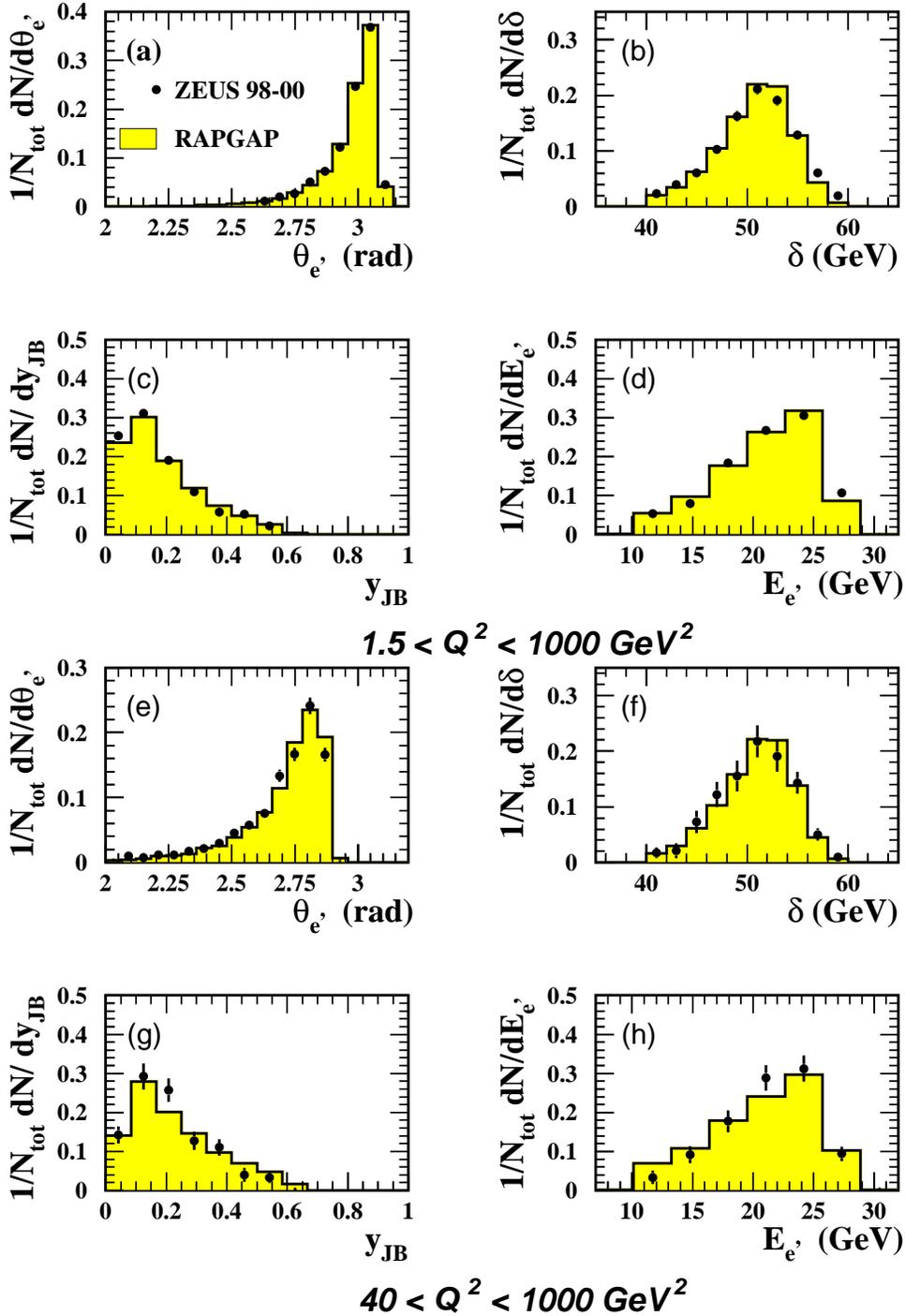,height=20.0cm}}
\caption[ig00_pap_ele]
{Reconstructed DIS variables for events with $D^*$ candidates (after background 
subtraction) for data (points) compared to detector-level {\sc Rapgap} predictions 
(shaded histograms): (a)-(d) show the distributions for $1.5<Q^2<1000$ GeV$^2$, 
while (e)-(h) are the same distributions but for $40<Q^2<1000$ GeV$^2$. All 
histograms are normalised to unit area.}
\label{ig00_pap_ele}
\end{center}
\end{figure}

Figure~\ref{ig00_pap_ele} shows distributions of DIS variables for $D^*$ events 
(after background subtraction) for data compared to detector-level {\sc Rapgap} 
predictions. The distributions, which are normalised to unit area, are shown 
separately for two $Q^2$ intervals: \mbox{$1.5<Q^2<1000$ GeV$^2$} and 
\mbox{$40<Q^2<1000$ GeV$^2$}. The {\sc Rapgap} predictions are in good 
agreement with the data distributions for both the scattered-lepton and hadronic 
variables. The description is similarly good for the two $Q^2$ ranges. This good 
description gives confidence in the use of the {\sc Rapgap} MC to 
correct the data for detector effects. The {\sc Herwig} MC gives a similarly 
good representation of the data (not shown) and is used to estimate the systematic 
uncertainty, arising from the model in the correction procedure, as described in 
Section~\ref{sec:syscheck}.

The cross sections for a given observable $Y$ were determined using 

\begin{equation}
\frac {d\sigma}{dY} = 
\frac {N } {A \cdot \mathcal {L} \cdot B \cdot \Delta Y},  
\nonumber 
\end{equation}

where $N$ is the number of $D^*$ events in a bin of size $\Delta Y$, $A$ is the 
acceptance (which takes into account migrations, efficiencies and QED radiative 
effects for that bin) and $\mathcal {L}$ is the integrated luminosity. The 
product, $B$, of the appropriate branching ratios for the $D^*$ and \dz\ was 
set to $(2.57\pm 0.06)\%$~\cite{pr:d66:010001}. 

\section{$D^*$ rates in  $e^-p$ and $e^+p$ interactions} 

The $D^*$ production rate, $r = N/\mathcal{L}$, in the $e^-p$ data set is 
systematically higher than that in the $e^+p$ data set. This difference 
increases with $Q^2$; for example, the  ratio of the rates, $r^{e^-p}/r^{e^+p}$, 
is equal to $1.12 \pm0.06$ for \mbox{$1.5<Q^2<1000$ GeV$^2$}, while for 
\mbox{$40<Q^2<1000$ GeV$^2$} it is $1.67\pm 0.21$ (only statistical errors are 
given). Such a difference in production cross sections is not expected from known 
physics processes.

A detailed study was performed to understand whether 
any instrumental effects could account for the difference between the two data 
sets. No such effect was seen in inclusive DIS where the ratio of $e^-p$ 
to $e^+p$ rates is consistent with unity. 
The rate for the wrong-charge background under the $D^*$ mass peak in 
$e^-p$ data agreed well with the wrong-charge rate in $e^+p$ data. For example, 
for $Q^2>40$ GeV$^2$, where the largest difference exists, the ratio of the 
rates for wrong-charge track combinations in $e^-p$ and $e^+p$ data is 
$0.95 \pm 0.09$. For both $e^-p$ and $e^+p$ interactions, the number of $D^{*+}$ 
mesons was consistent with the number of $D^{*-}$ for the entire $Q^2$ range 
studied.  Different reconstruction methods, cuts, background-subtraction methods 
and the time dependence of the 
difference were also investigated. None of these checks gave an 
indication of the source of the observed difference between the $D^*$ rates 
in $e^-p$ and $e^+p$ for $Q^2>40$ GeV$^2$. The cross sections were measured 
separately for $e^-p$ and $e^+p$ data and are discussed in 
Section~\ref{sec:xsec}. The difference in observed rate is assumed to be a 
statistical fluctuation and the two sets of data were combined for the 
final results.

\section{Experimental and theoretical uncertainties} 
\label{sec:syscheck}

\subsection{Experimental uncertainties}

The systematic uncertainties of the measured cross sections were determined by 
changing the selection cuts or the analysis procedure in turn and repeating the 
extraction of the cross sections~\cite{thesis:robins:2003}. The following 
systematic studies have been carried out (the resulting uncertainty on the 
total cross section is given in parentheses):   

\begin{itemize}

\item[$\bullet$]
event reconstruction and selection ($^{+2.3}_{-1.9}\%$). The following systematic 
checks were performed for this category:

\begin{itemize}

\item the cut on $y_e$ was changed to $y_{e}\> \leq\>  0.90$;

\item the cut on $y_{\mathrm{JB}}$ was changed to $y_{\mathrm{JB}}\> \geq\>  0.03$;

\item the cut on $\delta$ was changed to $42 \>  \leq \> \delta\>  \leq\>  57$ GeV;

\item the cut on the $|Z_{\rm vertex}|$ was changed to $|Z_{\rm vertex}| < 45$ cm;

\item the cut on $E_{e^\prime}$ was changed to $E_{e^\prime} > 11$~GeV;

\item the cut on the position of the scattered lepton in the RCAL was increased 
      by 1~cm;

\item the electron method was used, except for cases when the scattered-lepton 
      track was reconstructed by the CTD. In the latter case, the $\mathrm{DA}$ 
      method, which has the best resolution at high $Q^2$, was used;

\item the energy of the scattered electron was raised and lowered by $1\%$ in the 
      MC only, to account for the uncertainty in the CAL energy scale;

\item the energy of the hadronic system was raised and lowered by $3\%$ in the MC 
      only, to account for the uncertainty in the hadronic CAL energy scale;

\item the reconstructed SRTD hit position was shifted by $\pm 2$~mm to account for 
      the uncertainty in the SRTD-RCAL alignment.

\end{itemize}

\item[$\bullet$]
uncertainties related to the $D^*$ reconstruction ($^{+2.9}_{-1.6}\%$). The 
following systematic checks were performed for this category:

\begin{itemize}

\item tracks were required to have \mbox{$\mid \eta \mid < 1.75$}, in addition 
      to the requirement on the number of superlayers;

\item the cut on the minimum transverse momentum for the $\pi$ and $K$ candidates 
      was raised and lowered by 0.1~GeV;

\item the cut on the minimum transverse momentum for the $\pi_s$ was raised and 
      lowered by 0.02~GeV;

\item the signal region for the $M(D^0)$ was widened and narrowed symmetrically 
      around the centre by 0.01~GeV;

\item the signal region for the $\Delta M$ was widened symmetrically 
      around the centre by 0.003~GeV.

\end{itemize}

\item[$\bullet$]
the acceptance was determined using {\sc Herwig} instead of {\sc Rapgap} 
($-2.7\%$); 

\item[$\bullet$]
the uncertainty in the luminosity measurement (2.2\%).

\end{itemize}

The cross section obtained using the fit was in good agreement 
with that obtained by subtracting the background using the wrong-charge candidates. 
These estimations were also made in each bin in which the differential cross 
sections were measured. The overall systematic uncertainty was determined by 
adding the above uncertainties in quadrature. The normalisation uncertainties 
due to the luminosity-measurement error, and those due to the $D^*$ and $D^0$ 
decay branching ratios of 2.5\%~\cite{pr:d66:010001}, were not included in the 
systematic uncertainties for the differential cross sections. 

\subsection{Theoretical uncertainties}
\label{sec:theo_unc}

The NLO QCD predictions for $D^*$ production are affected by the systematic 
uncertainties listed below. Typical values for the systematic uncertainty are 
quoted for the total cross section:

\begin{itemize}

\item the proton PDF. The CTEQ5F3 and GRV98-HO~\cite{epj:c5:461} PDFs were used 
      to check  the sensitivity of the predictions to different parameterisations 
      of the gluon 
      density in the proton. The appropriate masses used in the fit to 
      determine the PDF were also used in HVQDIS, i.e. 1.3~GeV for CTEQ5F3 and 
      1.4~GeV for GRV98-HO. The change in the cross section was $+2.0\%$ using 
      CTEQ5F3 and $-16\%$ using GRV98-HO;

\item the mass of the charm quark $\left( ^{+9.7}_{-9.1}\% \right)$. The charm 
      mass was changed consistently in the PDF fit and in HVQDIS by 
      $\mp 0.15$~GeV. The largest effect was at low $p_T(D^*)$;

\item the renormalisation and factorisation scale, $\mu$ 
      $\left( ^{+4}_{-1}\% \right)$. The scale was changed by a factor of 0.5 
      and 2; another scale, 2$m_c$, was also used~\cite{pr:d57:2806}. The 
      maximum of $\sqrt{Q^2/4+m_c^2}$ and 2$m_c$ as a function of $Q^2$ was 
      taken as the scale to estimate the upward uncertainty;

\item the ZEUS PDF uncertainties propagated from the experimental 
      uncertainties of the fitted data ($\pm 5\%$). The change in the cross 
      section was independent of the kinematic region;

\item uncertainty in the fragmentation $\left( ^{+6}_{-4}\% \right)$. The 
      parameter $\epsilon$ in the Peterson fragmentation function was changed 
      by $\pm 0.015$.

\end{itemize}

The first source of systematic uncertainty is shown separately in the figures. 
The last four were added in quadrature and displayed as a band in the figures. 
An additional normalisation uncertainty of $3\%$~\cite{hep-ex-9912064} on the 
hadronisation fraction $f(c \to D^*)$ is not shown.

\section{Cross-section measurements}
\label{sec:xsec}

\subsection{Visible cross sections}

The overall acceptance after applying the selection criteria described in 
Sections~\ref{sub:ds} and~\ref{sub:dssel} for $1.5<Q^2<1000$ GeV$^2$, 
$0.02<y<0.7$, \mbox{$1.5<\ptds<15$ GeV} and $|\etads|<1.5$ calculated with 
{\sc Rapgap} is $31\%$, both for $e^-p$ and $e^+p$ data. The total 
cross sections in the same region are:

$$
\sigma(e^-p \rightarrow e^-D^{*}X) =
9.37 \pm 0.44(\mbox{stat.})^{+0.59}_{-0.52}(\mbox{syst.}) \pm 0.23(\mbox{BR}) \mbox{ nb};
$$

$$
\sigma(e^+p \rightarrow e^+D^{*}X) =
8.20 \pm 0.22(\mbox{stat.})^{+0.39}_{-0.36}(\mbox{syst.}) \pm 0.20(\mbox{BR})  \mbox{ nb}, 
$$

where the final uncertainty arises from the uncertainty on the branching 
ratios for the $D^*$ and $D^0$. The $D^*$ cross section for $e^+p$ data is 
consistent with the previously 
published result~\cite{epj:c12:35} obtained at a proton beam energy of 
820 GeV. According to 
HVQDIS, a $5\%$ increase in the $D^*$  cross section is expected when the 
proton energy increases from 820 to 920 GeV. 

The cross section obtained from the combined sample is:

$$
\sigma(e^\pm p \rightarrow e^\pm D^{*}X) =
8.44 \pm 0.20(\mbox{stat.})^{+0.37}_{-0.36}(\mbox{syst.})  \pm 0.21(\mbox{BR}) \mbox{ nb}. 
$$

The prediction from the HVQDIS program is \mbox{8.41 $^{+1.09}_{-0.95}$ nb}, 
in good agreement with the data. The uncertainty in the HVQDIS 
prediction arises from the sources discussed in Section~\ref{sec:theo_unc} (excluding 
that from using a different proton PDF) and 
is about 2.5 times the size of the uncertainty in the measurement. A 
contribution to the total cross sections arises from $D^*$ mesons produced in 
$b\bar{b}$ events. The $D^*$ cross section arising from $b\bar{b}$ production 
was estimated, as described in Section~\ref{sec:theory}, to be $0.17$ nb for 
\mbox{$Q^2>1.5$ GeV$^2$}. The measured differential cross 
sections include a component from beauty production. Therefore, all 
NLO predictions include a $b \bar{b}$ contribution calculated in each bin. 
For the extraction of $F_2^{c\bar{c}}$, the predicted value of 
$b \bar{b}$ production was subtracted from the data.

\subsection{Differential cross-section measurements}

\begin{figure}[htp]
\vspace{-2.0cm}
\begin{center}
\mbox{\epsfig{file=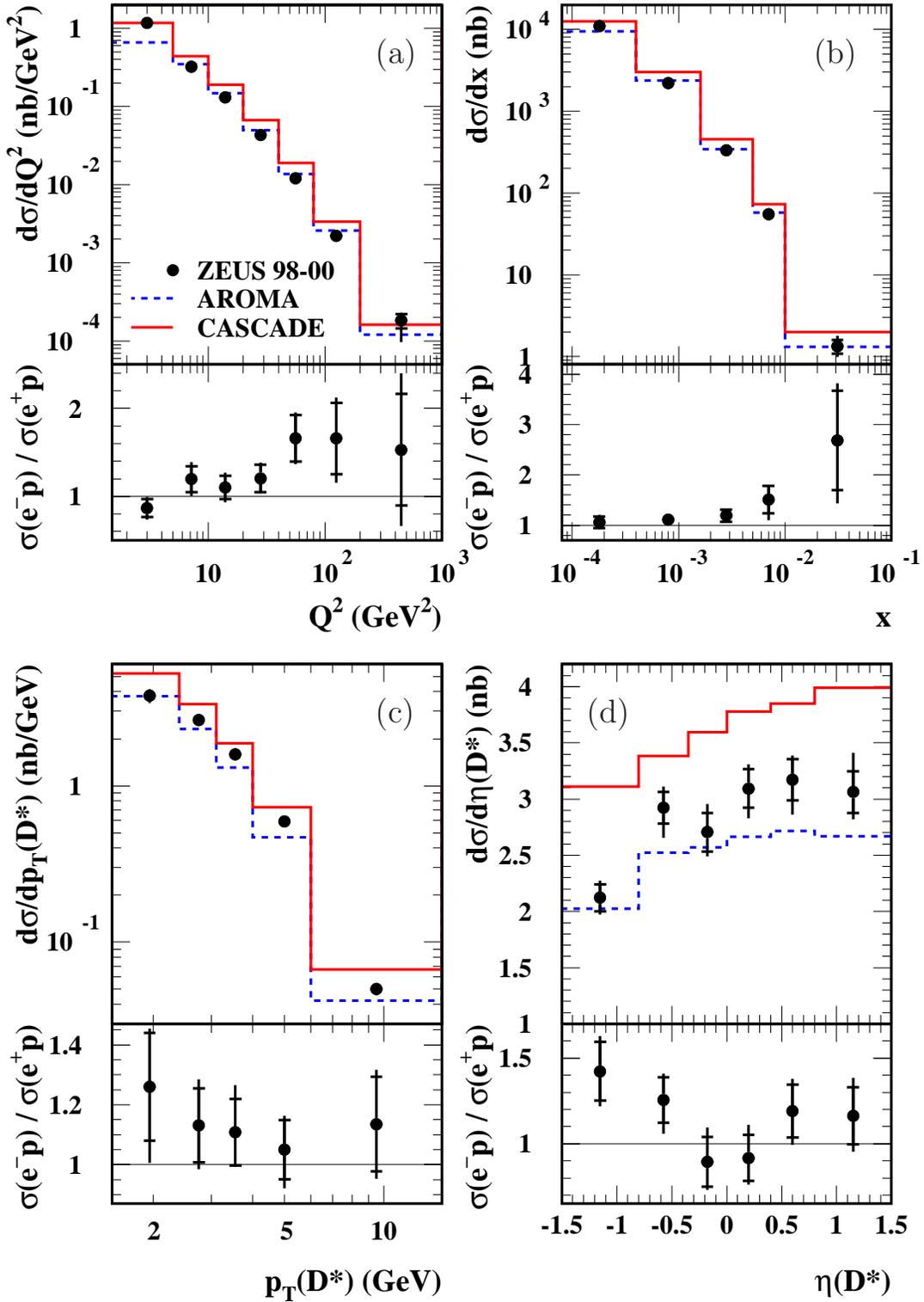,height=21.cm}}
\put(-240,556){\makebox(0,0)[tl]{\large (a)}}
\put(-47,556){\makebox(0,0)[tl]{\large (b)}}
\put(-240,266){\makebox(0,0)[tl]{\large (c)}}
\put(-147,266){\makebox(0,0)[tl]{\large (d)}}
\vspace{0cm}
\caption[ig1_pap_mc]
{Differential $D^*$ cross sections, for $e^-p$ and $e^+p$ data combined, as a 
function of (a) $Q^2$, (b) $x$, (c) $p_T(D^*)$ and (d) $\eta(D^*)$ compared with 
MC predictions. The inner error bars show the statistical uncertainties and the 
outer bars show the statistical and systematic uncertainties added in 
quadrature. Predictions from the {\sc Aroma} (dashed line) and {\sc Cascade} (solid line) 
MC programs are shown. The ratios of the cross sections for $e^-p$ and $e^+p$ 
data are also shown beneath each plot.}
\label{ig1_pap_mc}
\end{center}
\end{figure}

The differential $D^*$ cross sections as a function of $Q^2$, $x$, $p_T(D^*)$ 
and $\eta(D^*)$ for the combined $e^-p$ and $e^+p$ data samples are shown in 
Fig.~\ref{ig1_pap_mc} and given in Table~\ref{tab:sd_bins}. The cross 
sections in $Q^2$ and $x$ both fall by about four orders of magnitude in the 
measured region. The cross-section $d\sigma/dp_T(D^*)$ falls by two orders of 
magnitude with increasing $p_T(D^*)$. The cross-section $d\sigma/d\eta(D^*)$ 
rises with increasing $\eta(D^*)$. The ratio of the $e^-p$ and $e^+p$ cross 
sections, also shown in Fig.~\ref{ig1_pap_mc} and given in 
Table~\ref{tab:sd_bins}, tends to increase with increasing $Q^2$ and $x$. 
Neither the NLO calculations nor the MCs based on LO matrix elements and parton 
showers depend on the charge of the lepton in $ep$ interactions. 

The data in Fig.~\ref{ig1_pap_mc} are compared with predictions from the 
MC generators {\sc Aroma} and {\sc Cascade}. The prediction from {\sc Aroma} 
is generally below the data, particularly at low $Q^2$ and medium to high 
$p_T(D^*)$. In contrast, the prediction from {\sc Cascade} agrees at low $Q^2$, 
but generally lies above the data. Both MC predictions describe the shapes of 
the cross-sections $d\sigma/dx$ and $d\sigma/d\eta(D^*)$ reasonably well. The 
uncertainties in these MC predictions are difficult to estimate and may be 
large.

\begin{figure}
\vspace{-2.0cm}
\begin{center}
\mbox{\epsfig{file=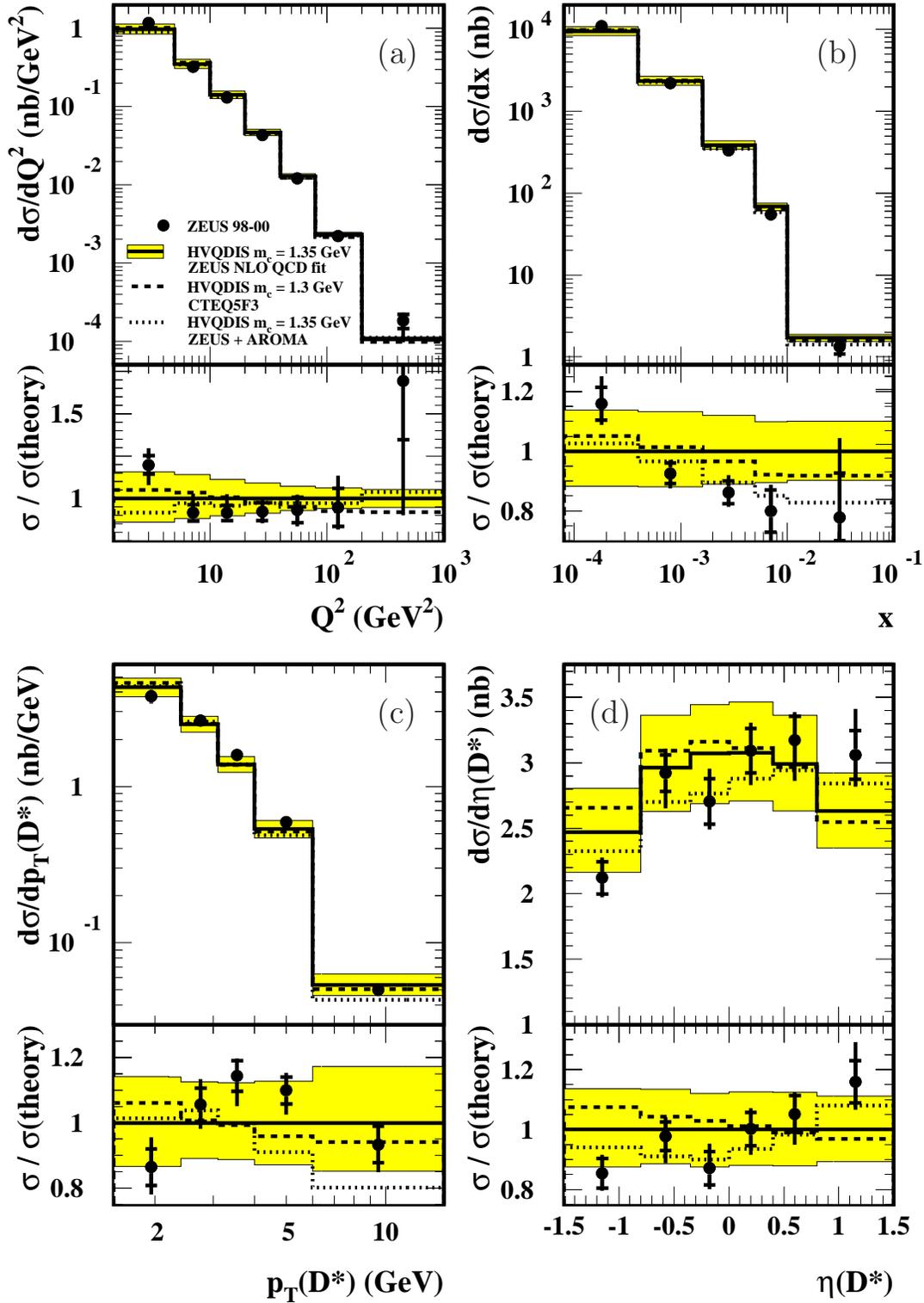,height=21.cm}}
\put(-240,556){\makebox(0,0)[tl]{\large (a)}}
\put(-47,556){\makebox(0,0)[tl]{\large (b)}}
\put(-240,266){\makebox(0,0)[tl]{\large (c)}}
\put(-147,266){\makebox(0,0)[tl]{\large (d)}}
\vspace{0cm}
\vspace{-0.5cm}
\caption[ig1_pap]
{Differential $D^*$ cross sections, for $e^-p$ and $e^+p$ data combined, as a 
function of (a) $Q^2$, (b) $x$, (c) $p_T(D^*)$ and (d) $\eta(D^*)$ compared to 
the NLO QCD calculation of HVQDIS. The inner error bars show the statistical 
uncertainties and the outer bars show the statistical and systematic 
uncertainties added in quadrature. Predictions from the ZEUS NLO QCD fit are 
shown for $m_c = 1.35$~GeV (solid line) with its associated uncertainty (shaded 
band) as discussed in the text. Predictions using the CTEQ5F3 PDF (dashed-dotted 
line) and an alternative hadronisation scheme (dotted line) are displayed. The 
ratios of the cross sections to the central HVQDIS prediction are also shown 
beneath each plot.}
\label{ig1_pap}
\end{center}
\end{figure}

In Fig.~\ref{ig1_pap}, the same data are compared with the NLO 
calculation implemented in the HVQDIS program. The predictions used the 
default parameter settings as discussed in Section~\ref{sec:theory}, with the 
uncertainties described in Section~\ref{sec:theo_unc}. 
Predictions using an alternate PDF, CTEQ5F3, and an alternate hadronisation 
scheme, from {\sc Aroma}, are also shown. The differences between the 
predictions, which are comparable to the uncertainties in the data, demonstrate 
the sensitivity of this measurement to the gluon distribution in the proton. 
The ratio of data to theory is displayed for each variable. For the cross 
sections as a function of $Q^2$ and $x$, the NLO predictions give a 
reasonable description of the data over four orders of magnitude in the 
cross section. For $d\sigma/dQ^2$, the description of the data is similar 
over the whole range in $Q^2$, even though HVQDIS is expected to be most 
accurate when $Q^2 \sim m_c^2$. The NLO calculation does, however, exhibit a 
somewhat different shape, particularly for $d\sigma/dx$, where the NLO is 
below the data at low $x$ and above the data at high $x$. The predictions 
using CTEQ5F3 instead of the ZEUS NLO fit, or using {\sc Aroma} for the 
hadronisation instead of the Peterson function, give better agreement with 
the data for the cross-section $d\sigma/dx$.  

The cross sections as a function of $p_T(D^*)$ and $\eta(D^*)$ are also 
reasonably well described by the NLO calculation. The prediction using the 
ZEUS NLO QCD fit gives a better description than that using CTEQ5F3 (and 
also better than the prediction using GRV98-HO, not shown), especially for 
the cross-section $d\sigma/d\eta(D^*)$. A better description of 
$d\sigma/d\eta(D^*)$ is also achieved~\cite{thesis:redondo:2001} by using 
{\sc Aroma} for the hadronisation, although, in this case, 
$d\sigma/dp_T(D^*)$ is not so well described. It should be noted that 
previous publications~\cite{epj:c12:35,pl:b528:199} revealed discrepancies 
in the forward $\eta(D^*)$ direction. This region can now be reasonably well 
described by a recent fit to the proton PDF as shown in Fig.~\ref{ig1_pap}(d). 
The data presented here are 
practically independent of the data used in the ZEUS NLO PDF fit to inclusive 
DIS data. Further refinement of NLO QCD fits and even the use of these data in 
future fits may achieve a better description.

\begin{figure}
\begin{center}
\mbox{\epsfig{file=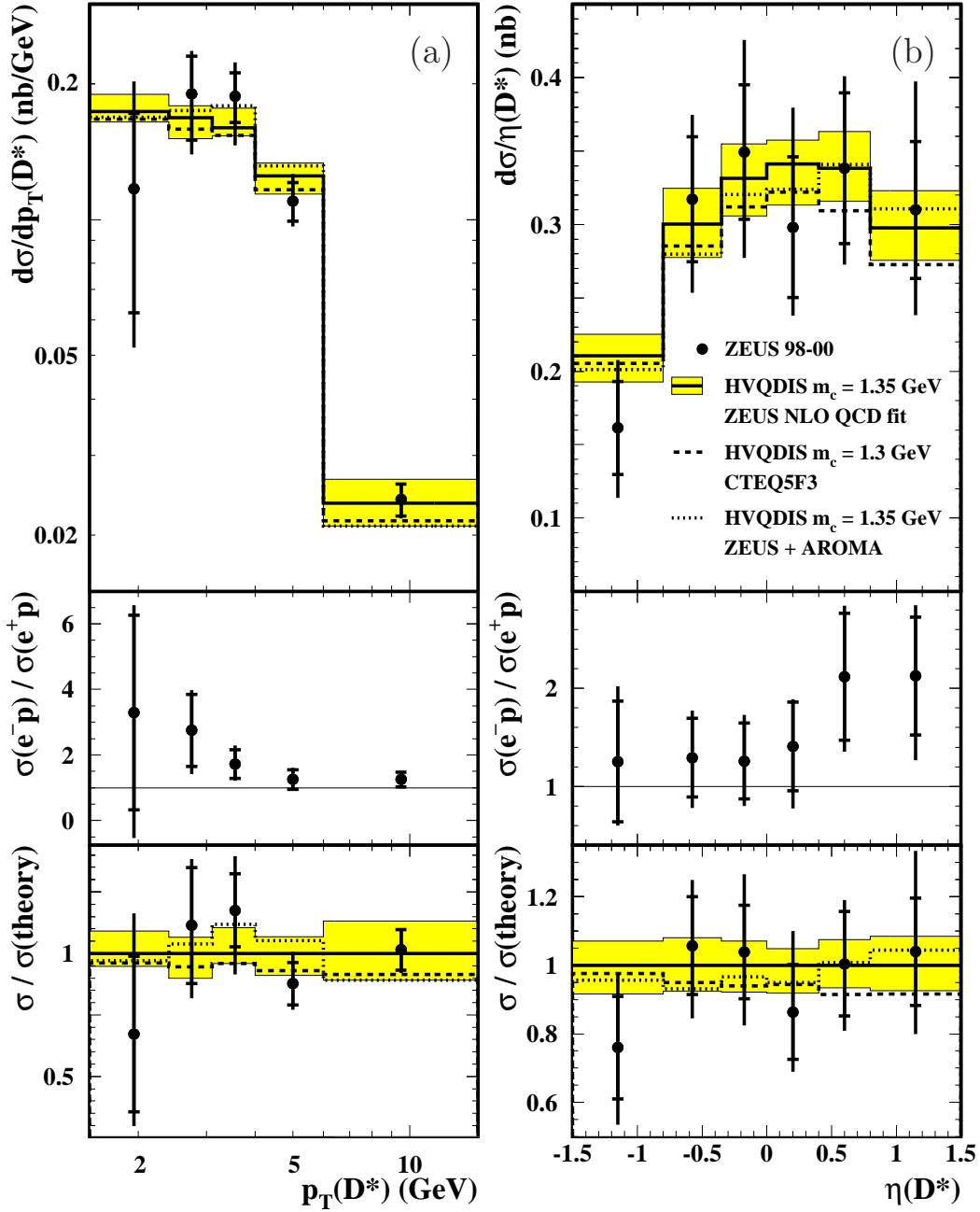,height=18.cm}}
\put(-230,476){\makebox(0,0)[tl]{\large (a)}}
\put(-35,476){\makebox(0,0)[tl]{\large (b)}}
\caption
{Differential $D^*$ cross sections, for $e^-p$ and $e^+p$ data combined, as a 
function of (a) $p_T(D^*)$ and (b) $\eta(D^*)$ for $Q^2 > 40$~GeV$^2$. The inner 
error bars show the statistical uncertainties and the outer bars show the 
statistical and systematic uncertainties added in quadrature. Predictions from 
the ZEUS NLO QCD fit are shown for $m_c = 1.35$~GeV (solid line) with its 
associated uncertainty (shaded band) as discussed in the text. Predictions using 
the CTEQ5F3 PDF (dashed-dotted line) and an alternative hadronisation scheme 
(dotted line) are displayed. The ratios of the cross sections for $e^-p$ and 
$e^+p$ data and for $e^-p$ and $e^+p$ data combined to the central HVQDIS 
prediction are also shown beneath each plot.} 
\label{see4_q2cut}
\end{center}
\end{figure}

Cross sections as a function of $\eta(D^*)$ and $p_T(D^*)$ were also measured 
for \mbox{$Q^2>40$ GeV$^2$}. The combined $e^-p$ and $e^+p$ data samples are 
given in Table~\ref{tab:q240_bins} and shown in Fig.~\ref{see4_q2cut} 
compared with the HVQDIS predictions. Although the HVQDIS calculation is not 
thought to be applicable at high $Q^2$, the data are well described. 
The high-$Q^2$ region is also where the difference in $e^-p$ and $e^+p$ data 
is most pronounced; the ratios of the cross sections are given in 
Table~\ref{tab:q240_bins}.

\section{Extraction of \ftwoccb}
\label{sec:extract}
 
The open-charm contribution, \ftwoccb, to the proton structure-function \ftwo\ 
can be defined in terms of the inclusive double-differential $c\bar{c}$ cross 
section in $x$ and \qsq\ by

\begin{equation}
\frac{d^2\sigma^{c\bar{c}} (x, Q^2)}{dxdQ^2} =
\frac{2\pi\alpha^2}{x Q^4}
\{ [1+(1-y)^2] F_2^{c\bar{c}}(x, Q^2) - y^2 F_L^{c\bar{c}}(x, Q^2) \} .
\label{eq:nc_charm}
\end{equation}

In this paper, the $c\bar{c}$ cross section is obtained by measuring the $D^*$ 
production cross section and employing the hadronisation fraction 
$f(c \rightarrow D^\ast)$ to derive the total charm cross section. Since only 
a limited kinematic region is accessible for the measurement of $D^*$ mesons, 
a prescription for extrapolating to the full kinematic phase space is needed. 

Since the structure function varies only slowly, it is assumed to be constant 
within a given \qsq\ and $y$ bin. Thus, the measured \ftwoccb\ in a bin $i$ 
is given by

\begin{equation}
F_{2,\rm meas}^{c\bar{c}}(x_i, Q^2_i) = \frac{\sigma_{i,\rm meas}(ep \rightarrow D^* X)}
                                      {\sigma_{i,\rm theo}(ep \rightarrow D^* X)}
                                      F_{2,\rm theo}^{c\bar{c}}(x_i, Q^2_i),
\end{equation}

where $\sigma_i$ are the cross sections in bin $i$ in the measured region of 
\ptds\ and $\eta(D^*)$. The value of $\ftwoccb_{\rm theo}$ was calculated from 
the NLO coefficient functions~\cite{pr:d67:012007}. The functional form of 
$\ftwoccb_{\rm theo}$ was used to quote the results for \ftwoccb\ at convenient 
values of $x_i$ and $ Q^2_i$ close to the centre-of-gravity of the bin. In this 
calculation, the same parton densities, charm mass ($m_c =$ 1.35 GeV), and 
factorisation and renormalisation scales ($\sqrt{4m_c^2 + \qsq}$) have been 
used as for the HVQDIS calculation of the differential cross sections. The 
hadronisation was performed using the Peterson fragmentation function.

The beauty contribution was subtracted from the data  using the theoretical 
prediction as described in Section~\ref{sec:theory}. At low $Q^2$ and high $x$, 
this fraction is small but it increases with increasing $Q^2$ and decreasing $x$. 
For the lower $x$ point at highest $Q^2$, the contribution from beauty 
production is about $7\%$ of that due to charm production. The contribution 
to the total cross section from 
$F_L^{c\bar{c}}$ calculated using the ZEUS NLO fit is, on average, $1.3\%$ and 
at most $4.7\%$ and is taken into account in the extraction of $F_2^{c\bar{c}}$. 
The size of the contribution from $F_L$ is similar to that in other PDFs.

Cross sections in the measured $D^*$ region and in the $Q^2$ and $y$ kinematic 
bins of Table~\ref{tab:dd_bins} were extrapolated to the full \ptds\ and \etads\ 
phase space using HVQDIS. These bins correspond to the $Q^2$ and $x$ values 
given in Table~\ref{tab:f2c_bins}, where the $F_2^{c\bar{c}}$ measurements are 
given. Typical extrapolation factors are between 4.7 at low \qsq and 1.5 at high 
$Q^2$, as in Table~\ref{tab:f2c_bins}. The following uncertainties 
of the extrapolation were evaluated:

\begin{itemize}

\item using the {\sc Aroma} fragmentation correction instead of the Peterson 
      fragmentation yielded changes of typically less than $10\%$ and not more 
      than 20\%. Although these values are not very significant compared to the 
      uncertainties in the data, the two corrections do produce a noticeable 
      change in the shape of the cross section as a function of $x$. The most 
      significant effects are in the highest $x$ bins for a given $Q^2$;

\item changing the charm mass by $\pm 0.15$~GeV consistently in the HVQDIS 
      calculation and in the calculation of $F_2^{c\bar{c}}$ leads to 
      differences in the extrapolation of $5\%$ at low $x$; the value 
      decreases rapidly to higher $x$; 

\item using the upper and lower predictions given by the uncertainty in the 
      ZEUS NLO PDF fit, propagated from the experimental uncertainties of the 
      fitted data, to perform the extraction of $F_2^{c\bar{c}}$ gives similar 
      values to the central measurement, with deviations typically less than 
      $1\%$; 

\item changing the contribution of beauty events subtracted from the data by 
      $^{+100}_{-50}\%$ gave 
      an uncertainty of typically $1-2\%$ and up to $8\%$ at low $x$ and 
      high $Q^2$.

\end{itemize}

These uncertainties were added in quadrature with the experimental systematic 
uncertainties when displayed in the figures and are given separately in 
Table~\ref{tab:f2c_bins}. Extrapolating the cross sections to the full $D^*$ 
phase space using the CTEQ5F3 proton PDF yielded differences compared to the 
ZEUS NLO QCD fit of less than $5\%$ for $Q^2>11$~GeV$^2$ and less than $10\%$ 
for $Q^2<11$~GeV$^2$. 

\begin{figure}
\vspace{-1.0cm}
\begin{center}
\mbox{\epsfig{file=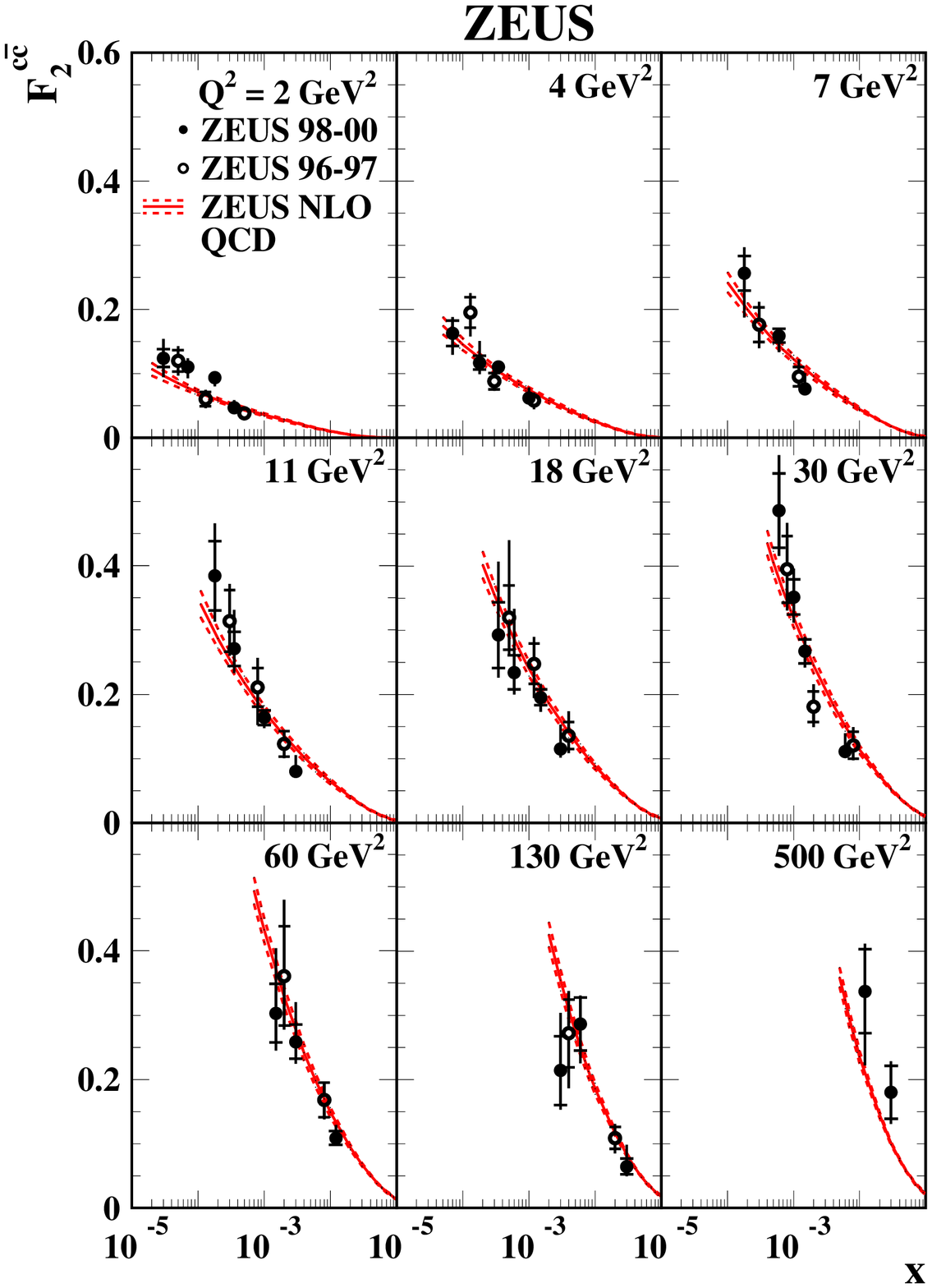,height=20cm}}
\caption
{The measured $F_2^{c \bar{c}}$ at $Q^2$ values between 2 and 500~GeV$^2$ as 
a function of $x$. The current data (solid points) are compared with the previous 
ZEUS measurement (open points). The data are shown with statistical uncertainties 
(inner bars) and statistical and systematic uncertainties added in quadrature 
(outer bars). The lower and upper curves show the fit uncertainty propagated from 
the experimental uncertainties of the fitted data.}
\label{f2charm_vs_x}
\end{center}
\end{figure}

The data are compared in Fig.~\ref{f2charm_vs_x} with the 
previous measurement~\cite{epj:c12:35} and with the ZEUS NLO QCD fit. The two 
sets of data are consistent\footnote{The first three points of the previous 
data were measured at $Q^2 = 1.8$~GeV$^2$ and not at 2~GeV$^2$, so they have 
been shifted to 2~GeV$^2$ using the ZEUS NLO QCD fit. All other points were 
measured at the same $Q^2$ values.}. The prediction describes the data well 
for all $Q^2$ and $x$ except for the lowest $Q^2$, where some difference is 
observed. The uncertainty on the theoretical prediction is that 
from the PDF fit propagated from the experimental uncertainties of the fitted 
data. At the lowest $Q^2$, the uncertainty in the data is comparable to the 
PDF uncertainty shown. This implies that the double-differential cross sections 
given in Table~\ref{tab:dd_bins} could be used as an additional constraint on 
the gluon density in the proton.

\begin{figure}
\vspace{-1.0cm}
\begin{center}
\mbox{\epsfig{file=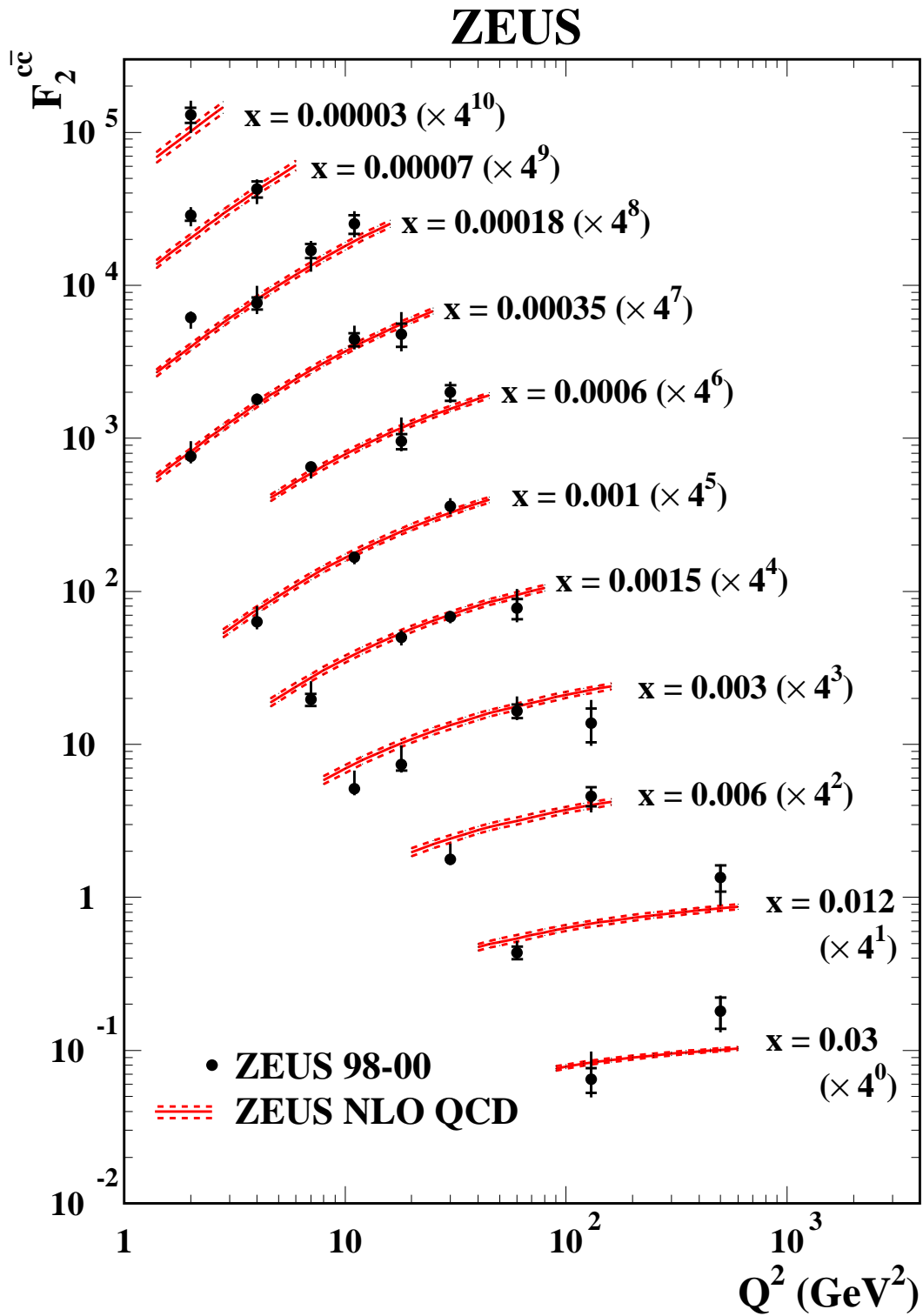,height=20cm}}
\caption
{The measured $F_2^{c \bar{c}}$ at $x$ values between 0.00003 and 0.03 as a 
function of $Q^2$. The data are shown with statistical uncertainties (inner bars) 
and statistical and systematic uncertainties added in quadrature (outer bars). 
The lower and upper curves show the fit uncertainty propagated from the 
experimental uncertainties of the fitted data.}
\label{f2charm_vs_q2}
\end{center}
\end{figure}

The values of \ftwoccb\ are presented as a function of $Q^2$ at fixed values of 
$x$ and compared with the ZEUS NLO QCD fit in Fig.~\ref{f2charm_vs_q2}. The 
data rise with increasing $Q^2$, with the rise becoming steeper at lower $x$, 
demonstrating the property of scaling violation in charm production. The data 
are well described by the prediction. 

\begin{figure}
\vspace{-1.0cm}
\begin{center}
\mbox{\epsfig{file=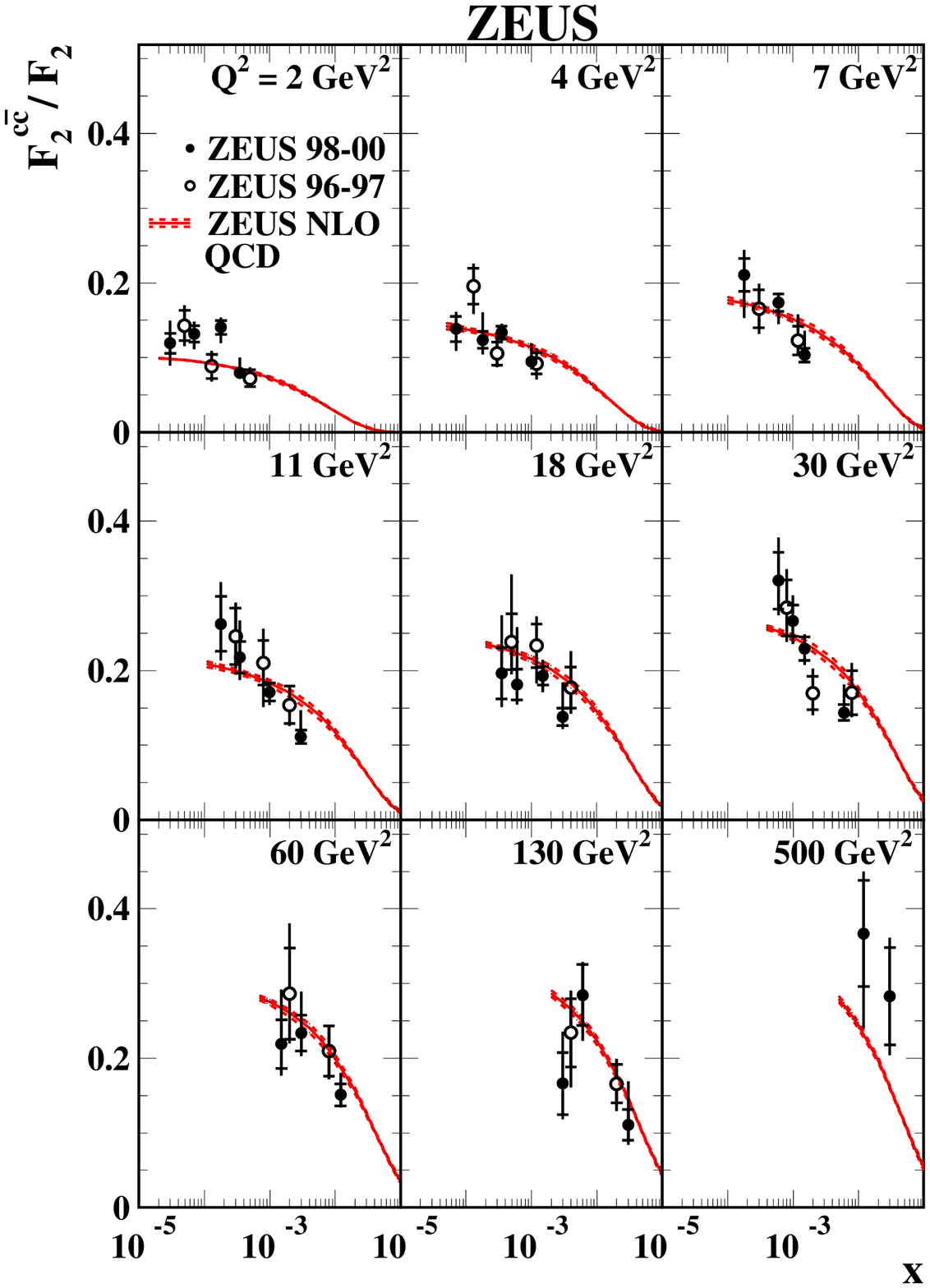,height=20cm}}
\caption
{The measured  ratio $F_2^{c \bar{c}}/F_2$ at $Q^2$ values between 2 and 
500~GeV$^2$ as a function of $x$. The current data (solid points) are compared 
with the previous ZEUS measurement (open points). The data are shown with 
statistical uncertainties (inner bars) and statistical and systematic 
uncertainties added in quadrature (outer bars). The lower and upper curves 
show the fit uncertainty propagated from the experimental uncertainties of the 
fitted data.}
\label{f2charm_over_f2}
\end{center}
\end{figure}

Figure~\ref{f2charm_over_f2} shows the ratio $F_2^{c\bar{c}}/F_2$ as a function 
of $x$ at fixed values of $Q^2$. The values of $F_2$ used to determine the ratio 
were taken from the ZEUS NLO QCD fit at the same values of $Q^2$ and $x$ at 
which \ftwoccb\ is quoted, and are given in Table~\ref{tab:f2c_bins}. The ratio 
\ftwoccb/\ftwo\ rises from $10\%$ to $30\%$ as $Q^2$ increases and $x$ decreases.
 
\section{Conclusions}

The production of $D^*$ mesons has been measured in deep inelastic scattering at 
HERA in the kinematic region $1.5 < Q^2 < 1000$~GeV$^2$, $0.02<y<0.7$, 
$1.5 < p_T(D^*) < 15$~GeV and $|\eta(D^*)|<1.5$. The data extend the previous 
analysis to higher $Q^2$ and have increased precision. 

Predictions from the {\sc Aroma} MC underestimate, and those from the 
{\sc Cascade} MC overestimate, the measured cross sections. 
Predictions from NLO QCD are in reasonable agreement with the measured cross 
sections, which show sensitivity to the choice of PDF and hence the gluon 
distribution in the proton. The ZEUS NLO PDF, which was fit to 
recent inclusive DIS data, gives the best description of the $D^*$ data. 
In particular, this is seen in the cross-section $d\sigma/d\eta(D^*)$. 

The double-differential cross section in $y$ and $Q^2$ has been 
measured and used to extract the open-charm contribution to $F_2$, by 
using the NLO QCD calculation to extrapolate outside the 
measured $p_T(D^*)$ and $\eta(D^*)$ region. Since, at low $Q^2$, 
the uncertainties of the data are comparable to those from the PDF fit,  
the measured differential cross sections in $y$ and $Q^2$ should be used 
in future fits to constrain the gluon density.

\section*{Acknowledgments}
The strong support and encouragement of the DESY Directorate have been 
invaluable, and we are much indebted to the HERA machine group for their 
inventiveness and diligent efforts. The design, construction and installation 
of the ZEUS detector have been made possible by the ingenuity and dedicated 
efforts of many people from inside DESY and from the home institutes who are 
not listed as authors. Their contributions are acknowledged with great 
appreciation. We thank J.~Collins, B.~Harris, G.~Ingelman and R.~Thorne for 
useful discussions. We also thank A.~Donnachie, P.V.~Landshoff, 
K.~Golec-Biernat, A.V.~Kiselev, V.A.~Petrov, A.V.~Berezhnoy and A.K.~Likhoded 
for providing us with their predictions.

\vfill\eject


{
\def\bibname{\Large\bf References}
\def\refname{\Large\bf References}
\pagestyle{plain}
\ifzeusbst
  \bibliographystyle{./BiBTeX/bst/l4z_default}
\fi
\ifzdrftbst
  \bibliographystyle{./BiBTeX/bst/l4z_draft}
\fi
\ifzbstepj
  \bibliographystyle{./BiBTeX/bst/l4z_epj}
\fi
\ifzbstnp
  \bibliographystyle{./BiBTeX/bst/l4z_np}
\fi
\ifzbstpl
  \bibliographystyle{./BiBTeX/bst/l4z_pl}
\fi
{\raggedright
\bibliography{./BiBTeX/user/syn.bib,%
              ./BiBTeX/bib/l4z_articles.bib,%
              ./BiBTeX/bib/l4z_books.bib,%
              ./BiBTeX/bib/l4z_conferences.bib,%
              ./BiBTeX/bib/l4z_h1.bib,%
              ./BiBTeX/bib/l4z_misc.bib,%
              ./BiBTeX/bib/l4z_old.bib,%
              ./BiBTeX/bib/l4z_preprints.bib,%
              ./BiBTeX/bib/l4z_replaced.bib,%
              ./BiBTeX/bib/l4z_temporary.bib,%
              ./BiBTeX/bib/l4z_zeus.bib}}
}
\vfill\eject


\begin{table}[p]
\centering
\begin{tabular}{|c|ccc|c|}
\hline 
$Q^2$ bin ($\gev^2$) & $d\sigma / dQ^2$ & $\Delta_{\rm stat}$ & $\Delta_{\rm syst}$ & $\sigma(e^-p)/\sigma(e^+p)$ \\ 
& & (nb/$\gev^2$)& & \\
\hline
1.5, 5    & $1.18$    & $\pm 0.05$    &   $ ^{+0.08} _{-0.10}$       & $0.86 \pm 0.10 ^{+0.08}_{-0.08}$\\
5, 10     & $0.323$   & $\pm 0.017$   &   $ ^{+0.037} _{-0.010}$     & $1.20 \pm 0.15 ^{+0.13}_{-0.13}$\\
10, 20    & $0.130$   & $\pm 0.007$   &   $ ^{+0.014} _{-0.003}$     & $1.10 \pm 0.13 ^{+0.11}_{-0.11}$\\
20, 40    & $0.044$   & $\pm 0.002$   &   $ ^{+0.003} _{-0.002}$     & $1.20 \pm 0.16 ^{+0.09}_{-0.07}$\\
40, 80    & $0.012$   & $\pm 0.001$   &   $ ^{+0.001} _{-0.001}$     & $1.66 \pm 0.26 ^{+0.13}_{-0.14}$\\
80, 200   & $0.0022$  & $\pm 0.0003$  &   $ ^{+0.0003} _{-0.0001}$   & $1.66 \pm 0.41 ^{+0.22}_{-0.30}$\\
200, 1000 & $0.00018$ & $\pm 0.00004$ &   $ ^{+0.00003} _{-0.00008}$ & $1.53 \pm 0.64 ^{+0.56}_{-0.59}$\\
\hline 
\hline
$x$ bin & $d\sigma / dx$ & $\Delta_{\rm stat}$ & $\Delta_{\rm syst}$ & $\sigma(e^-p)/\sigma(e^+p)$ \\
& & (nb) & & \\
\hline
0.00008, 0.0004 & $11035$ & $\pm 524$  & $ ^{+716} _{-420}$   & $1.06 \pm 0.12 ^{+0.08}_{-0.07}$\\
0.0004, 0.0016  & $2193$  & $\pm 81.8$ & $ ^{+73.2} _{-89.1}$ & $1.11 \pm 0.10 ^{+0.07}_{-0.07}$\\
0.0016, 0.005   & $335$   & $\pm 15.0$ & $ ^{+16.6} _{-11.5}$ & $1.19 \pm 0.12 ^{+0.08}_{-0.06}$\\
0.005, 0.01     & $54.9$  & $\pm 4.9$  & $ ^{+3.7} _{-7.3}$   & $1.51 \pm 0.27 ^{+0.09}_{-0.31}$\\
0.01, 0.1       & $1.34$  & $\pm 0.26$ & $ ^{+0.38} _{-0.22}$ & $2.69 \pm 0.99 ^{+0.56}_{-0.76}$\\
\hline
\hline
$p_T(D^*)$ bin ($\gev$) & $d\sigma / dp_T(D^*)$ & $\Delta_{\rm stat}$ & $\Delta_{\rm syst}$ & $\sigma(e^-p)/\sigma(e^+p)$ \\
& & (nb/$\gev$) & & \\
\hline
1.5, 2.4  & $3.76$  & $ \pm 0.24$  & $ ^{+0.31} _{-0.27}$   & $1.26 \pm 0.18 ^{+0.07}_{-0.18}$\\
2.4, 3.1  & $2.64$  & $ \pm 0.13$  & $ ^{+0.15} _{-0.13}$   & $1.13 \pm 0.12 ^{+0.09}_{-0.08}$\\
3.1, 4.0  & $1.60$  & $ \pm 0.07$  & $ ^{+0.04} _{-0.11}$   & $1.11 \pm 0.11 ^{+0.11}_{-0.03}$\\
4.0, 6.0  & $0.59$  & $ \pm 0.02$  & $ ^{+0.02} _{-0.03}$   & $1.05 \pm 0.10 ^{+0.06}_{-0.08}$\\
6.0, 15   & $0.050$ & $ \pm 0.003$ & $ ^{+0.002} _{-0.003}$ & $1.14 \pm 0.16 ^{+0.09}_{-0.09}$\\
\hline
\hline
$\eta(D^*)$ bin & $d\sigma / d\eta(D^*)$ & $\Delta_{\rm stat}$ & $\Delta_{\rm syst}$ & $\sigma(e^-p)/\sigma(e^+p)$ \\
& & (nb) & & \\
\hline
$-1.5,$ $-0.8$  & $2.12$ & $ \pm 0.12$ & $ ^{+0.09} _{-0.08}$ & $1.42 \pm 0.17 ^{+0.11}_{-0.11}$\\
$-0.8,$ $-0.35$ & $2.92$ & $ \pm 0.14$ & $ ^{+0.13} _{-0.23}$ & $1.26 \pm 0.13 ^{+0.08}_{-0.15}$\\
$-0.35,$ 0.0    & $2.71$ & $ \pm 0.17$ & $ ^{+0.18} _{-0.13}$ & $0.89 \pm 0.15 ^{+0.14}_{-0.07}$\\
 0.0, 0.4       & $3.09$ & $ \pm 0.17$ & $ ^{+0.13} _{-0.20}$ & $0.92 \pm 0.14 ^{+0.14}_{-0.08}$\\
 0.4, 0.8       & $3.17$ & $ \pm 0.18$ & $ ^{+0.11} _{-0.25}$ & $1.19 \pm 0.16 ^{+0.11}_{-0.12}$\\
 0.8, 1.5       & $3.06$ & $ \pm 0.19$ & $ ^{+0.29} _{-0.16}$ & $1.16 \pm 0.17 ^{+0.15}_{-0.13}$\\
\hline
\end{tabular}
\caption{\textit{Measured differential cross sections as a function of $Q^2$, $x$, 
$p_T(D^*)$ and $\eta(D^*)$ for $1.5<Q^2<1000$ GeV$^2$, $0.02<y<0.7$, 
\mbox{$1.5<\ptds<15$ GeV} and $|\etads|<1.5$. The statistical and systematic 
uncertainties are shown separately. The ratio of the cross sections for $e^-p$ and 
$e^+p$ data are also given with statistical and systematic uncertainties shown 
separately.}}
\label{tab:sd_bins}
\end{table}

\begin{table}[p]
\centering
\begin{tabular}{|c|ccc|c|}
\hline 
$p_T(D^*)$ bin ($\gev$) & $d\sigma / dp_T(D^*)$ & $\Delta_{\rm stat}$ & $\Delta_{\rm syst}$ & $\sigma(e^-p)/\sigma(e^+p)$ \\
& & (nb/$\gev$) & & \\
\hline
$1.5, 2.4$ & $0.117$ & $ \pm 0.055$ & $ ^{+0.065} _{-0.035}$ & $3.29 \pm 2.97 ^{+1.39}_{-2.41}  $\\
$2.4, 3.1$ & $0.190$ & $ \pm 0.040$ & $ ^{+0.023} _{-0.031}$ & $2.75 \pm 1.10 ^{+0.55}_{-0.76}  $\\
$3.1, 4.0$ & $0.188$ & $ \pm 0.024$ & $ ^{+0.026} _{-0.034}$ & $1.72 \pm 0.44 ^{+0.37}_{-0.26}  $\\
$4.0, 6.0$ & $0.110$ & $ \pm 0.011$ & $ ^{+0.012} _{-0.008}$ & $1.25 \pm 0.30 ^{+0.20}_{-0.13}  $\\
$6.0, 15$  & $0.024$ & $ \pm 0.002$ & $ ^{+0.001} _{-0.001}$ & $1.25 \pm 0.23 ^{+0.07}_{-0.05}  $\\
\hline
\hline
$\eta(D^*)$ bin & $d\sigma / d\eta(D^*)$ & $\Delta_{\rm stat}$ & $\Delta_{\rm syst}$ & $\sigma(e^-p)/\sigma(e^+p)$ \\
& & (nb) & & \\
\hline
$-1.5,$ $-0.8$  & $0.161$ & $ \pm 0.032$ & $ ^{+0.033} _{-0.036}$ & $1.25 \pm 0.62 ^{+0.46}_{-0.22}  $\\
$-0.8,$ $-0.35$ & $0.317$ & $ \pm 0.043$ & $ ^{+0.039} _{-0.047}$ & $1.29 \pm 0.40 ^{+0.26}_{-0.32}  $\\
$-0.35,$ 0.0    & $0.349$ & $ \pm 0.046$ & $ ^{+0.061} _{-0.056}$ & $1.26 \pm 0.39 ^{+0.27}_{-0.24}  $\\
 0.0, 0.4       & $0.298$ & $ \pm 0.048$ & $ ^{+0.066} _{-0.036}$ & $1.41 \pm 0.45 ^{+0.16}_{-0.44}  $\\
 0.4, 0.8       & $0.338$ & $ \pm 0.051$ & $ ^{+0.036} _{-0.041}$ & $2.12 \pm 0.65 ^{+0.33}_{-0.41}  $\\
 0.8, 1.5       & $0.310$ & $ \pm 0.047$ & $ ^{+0.074} _{-0.054}$ & $2.13 \pm 0.60 ^{+0.40}_{-0.62}  $\\
\hline
\end{tabular}
\caption{\textit{Measured differential cross sections as a function of $Q^2$, $x$, 
$p_T(D^*)$ and $\eta(D^*)$ for $40<Q^2<1000$ GeV$^2$, $0.02<y<0.7$, 
\mbox{$1.5<\ptds<15$ GeV} and $|\etads|<1.5$. The statistical and systematic 
uncertainties are shown separately.}}
\label{tab:q240_bins}
\end{table}

\begin{table}[htb]
\begin{center}
\begin{tabular}{|l|c|cccc|c|}
\hline
$Q^2$ bin $(\gev^2)$ & $y$ bin  & $\sigma$ & $\Delta_{\rm stat}$ & $\Delta_{\rm syst}$ & (nb) & $\sigma_{\rm theo}^{b\bar{b}} (D^*)$ (nb) \\
\hline \hline
1.5, 3.5  & 0.70,  0.33 & $0.655$ & $ \pm 0.073$ & $ ^{+0.128} _{-0.100} $ & & 0.010 \\
          & 0.33,  0.18 & $0.842$ & $ \pm 0.070$ & $ ^{+0.066} _{-0.082} $ & & 0.008 \\
          & 0.18,  0.09 & $0.974$ & $ \pm 0.064$ & $ ^{+0.058} _{-0.117} $ & & 0.006 \\
          & 0.09,  0.02 & $0.648$ & $ \pm 0.048$ & $ ^{+0.095} _{-0.040} $ & & 0.002 \\
\hline
3.5, 6.5  & 0.70,  0.33 & $0.340$ & $ \pm 0.041$ & $ ^{+0.025} _{-0.032} $ & & 0.007 \\
          & 0.33,  0.18 & $0.379$ & $ \pm 0.034$ & $ ^{+0.103} _{-0.030} $ & & 0.006 \\
          & 0.18,  0.08 & $0.527$ & $ \pm 0.034$ & $ ^{+0.027} _{-0.021} $ & & 0.004 \\
          & 0.08,  0.02 & $0.365$ & $ \pm 0.025$ & $ ^{+0.036} _{-0.030} $ & & 0.001 \\
\hline
6.5, 9.0  & 0.70,  0.25 & $0.301$ & $ \pm 0.031$ & $ ^{+0.030} _{-0.065} $ & & 0.005 \\
          & 0.25,  0.08 & $0.384$ & $ \pm 0.025$ & $ ^{+0.008} _{-0.055} $ & & 0.004 \\
          & 0.08,  0.02 & $0.156$ & $ \pm 0.014$ & $ ^{+0.017} _{-0.009} $ & & 0.001 \\
\hline
9.0, 14   & 0.70,  0.35 & $0.225$ & $ \pm 0.031$ & $ ^{+0.032} _{-0.015} $ & & 0.005 \\
          & 0.35,  0.20 & $0.240$ & $ \pm 0.023$ & $ ^{+0.047} _{-0.019} $ & & 0.004 \\
          & 0.20,  0.08 & $0.314$ & $ \pm 0.022$ & $ ^{+0.002} _{-0.021} $ & & 0.003 \\
          & 0.08,  0.02 & $0.180$ & $ \pm 0.015$ & $ ^{+0.014} _{-0.007} $ & & 0.001 \\
\hline
14, 22    & 0.70,  0.35 & $0.130$ & $ \pm 0.022$ & $ ^{+0.043} _{-0.014} $ & & 0.004 \\
          & 0.35,  0.20 & $0.155$ & $ \pm 0.017$ & $ ^{+0.061} _{-0.012} $ & & 0.003 \\
          & 0.20,  0.08 & $0.263$ & $ \pm 0.016$ & $ ^{+0.022} _{-0.024} $ & & 0.003 \\
          & 0.08,  0.02 & $0.150$ & $ \pm 0.013$ & $ ^{+0.008} _{-0.012} $ & & 0.001 \\
\hline
22, 44    & 0.70,  0.35 & $0.226$ & $ \pm 0.026$ & $ ^{+0.027} _{-0.013} $ & & 0.006 \\
          & 0.35,  0.22 & $0.193$ & $ \pm 0.015$ & $ ^{+0.018} _{-0.015} $ & & 0.004 \\
          & 0.22,  0.08 & $0.261$ & $ \pm 0.018$ & $ ^{+0.010} _{-0.016} $ & & 0.004 \\
          & 0.08,  0.02 & $0.182$ & $ \pm 0.013$ & $ ^{+0.024} _{-0.005} $ & & 0.002 \\
\hline
44, 90    & 0.70,  0.28 & $0.141$ & $ \pm 0.020$ & $ ^{+0.040} _{-0.015} $ & & 0.006 \\
          & 0.28,  0.14 & $0.133$ & $ \pm 0.013$ & $ ^{+0.028} _{-0.010} $ & & 0.004 \\
          & 0.14,  0.02 & $0.130$ & $ \pm 0.013$ & $ ^{+0.010} _{-0.006} $ & & 0.003 \\
\hline
90, 200   & 0.70,  0.28 & $0.060$ & $ \pm 0.014$ & $ ^{+0.019} _{-0.006} $ & & 0.005 \\
          & 0.28,  0.14 & $0.076$ & $ \pm 0.011$ & $ ^{+0.003} _{-0.011} $ & & 0.003 \\
          & 0.14,  0.02 & $0.044$ & $ \pm 0.008$ & $ ^{+0.020} _{-0.006} $ & & 0.001 \\
\hline
200, 1000 & 0.70,  0.23 & $0.087$ & $ \pm 0.016$ & $ ^{+0.007} _{-0.023} $ & & 0.004 \\
          & 0.23,  0.02 & $0.050$ & $ \pm 0.011$ & $ ^{+0.006} _{-0.007} $ & & 0.001 \\
\hline
\end{tabular}
\end{center}
\caption{\textit{Measured cross sections in each of the $Q^2$ and $y$ bins for 
\mbox{$1.5<Q^2<1000$ GeV$^2$}, $0.02<y<0.7$, \mbox{$1.5<\ptds<15$ GeV} and 
\mbox{$|\etads|<1.5$.} The statistical and systematic uncertainties are shown separately. 
The prediction for the $\sigma_{\rm theo}^{b\bar{b}} (D^*)$ contribution from HVQDIS, which 
was subtracted from the data in the extraction of $F_2^{c\bar{c}}$,  is also shown.}}
\label{tab:dd_bins}
\end{table}

\begin{table}[htb]
\begin{center}

\begin{tabular}{|c|c|cccc|c|c|}
\hline
$Q^2$ ($\gev^2$) & $x$ & $F_2^{c\bar{c}}$ & $\Delta_{\rm stat}$ & $\Delta_{\rm syst}$ 
& $\Delta_{\rm theo}$ & extrapolation factor & $F_2$ \\
\hline \hline
2   & 0.00003 & $0.124$ & $ \pm 0.014$ & $ ^{+0.025} _{-0.019}$ & $ ^{+0.009} _{-0.017} $ & 4.17 & 0.983 \\
    & 0.00007 & $0.110$ & $ \pm 0.009$ & $ ^{+0.009} _{-0.011}$ & $ ^{+0.005} _{-0.009} $ & 3.02 & 0.817 \\
    & 0.00018 & $0.094$ & $ \pm 0.006$ & $ ^{+0.006} _{-0.011}$ & $ ^{+0.003} _{-0.006} $ & 3.07 & 0.672 \\
    & 0.00035 & $0.046$ & $ \pm 0.003$ & $ ^{+0.007} _{-0.003}$ & $ ^{+0.009} _{-0.000} $ & 4.72 & 0.591 \\
\hline
4   & 0.00007 & $0.163$ & $ \pm 0.020$ & $ ^{+0.012} _{-0.016}$ & $ ^{+0.011} _{-0.022} $ & 3.84 & 1.140 \\
    & 0.00018 & $0.117$ & $ \pm 0.011$ & $ ^{+0.032} _{-0.009}$ & $ ^{+0.005} _{-0.011} $ & 2.68 & 0.930 \\
    & 0.00035 & $0.110$ & $ \pm 0.007$ & $ ^{+0.006} _{-0.004}$ & $ ^{+0.003} _{-0.005} $ & 2.67 & 0.808 \\
    & 0.00100 & $0.062$ & $ \pm 0.004$ & $ ^{+0.006} _{-0.005}$ & $ ^{+0.015} _{-0.000} $ & 3.93 & 0.652 \\
\hline
7   & 0.00018 & $0.257$ & $ \pm 0.027$ & $ ^{+0.026} _{-0.057}$ & $ ^{+0.014} _{-0.028} $ & 3.18 & 1.195 \\
    & 0.00060 & $0.159$ & $ \pm 0.011$ & $ ^{+0.003} _{-0.023}$ & $ ^{+0.004} _{-0.006} $ & 2.34 & 0.907 \\
    & 0.00150 & $0.077$ & $ \pm 0.007$ & $ ^{+0.008} _{-0.004}$ & $ ^{+0.021} _{-0.000} $ & 3.31 & 0.737 \\
\hline
11  & 0.00018 & $0.384$ & $ \pm 0.054$ & $ ^{+0.056} _{-0.027}$ & $ ^{+0.025} _{-0.004} $ & 3.29 & 1.447 \\
    & 0.00035 & $0.271$ & $ \pm 0.027$ & $ ^{+0.054} _{-0.022}$ & $ ^{+0.009} _{-0.015} $ & 2.21 & 1.229 \\
    & 0.00100 & $0.164$ & $ \pm 0.012$ & $ ^{+0.001} _{-0.011}$ & $ ^{+0.003} _{-0.004} $ & 2.11 & 0.948 \\
    & 0.00300 & $0.080$ & $ \pm 0.007$ & $ ^{+0.006} _{-0.003}$ & $ ^{+0.024} _{-0.002} $ & 2.95 & 0.724 \\
\hline
18  & 0.00035 & $0.293$ & $ \pm 0.051$ & $ ^{+0.101} _{-0.032}$ & $ ^{+0.019} _{-0.028} $ & 2.96 & 1.476 \\
    & 0.00060 & $0.234$ & $ \pm 0.027$ & $ ^{+0.095} _{-0.018}$ & $ ^{+0.009} _{-0.012} $ & 1.94 & 1.280 \\
    & 0.00150 & $0.196$ & $ \pm 0.012$ & $ ^{+0.017} _{-0.018}$ & $ ^{+0.005} _{-0.003} $ & 1.90 & 1.001 \\
    & 0.00300 & $0.115$ & $ \pm 0.010$ & $ ^{+0.006} _{-0.009}$ & $ ^{+0.036} _{-0.001} $ & 2.69 & 0.831 \\
\hline
30  & 0.00060 & $0.487$ & $ \pm 0.058$ & $ ^{+0.059} _{-0.029}$ & $ ^{+0.026} _{-0.029} $ & 2.47 & 1.510 \\
    & 0.00100 & $0.352$ & $ \pm 0.027$ & $ ^{+0.033} _{-0.027}$ & $ ^{+0.011} _{-0.010} $ & 1.70 & 1.303 \\
    & 0.00150 & $0.267$ & $ \pm 0.019$ & $ ^{+0.010} _{-0.017}$ & $ ^{+0.007} _{-0.005} $ & 1.69 & 1.160 \\
    & 0.00600 & $0.111$ & $ \pm 0.008$ & $ ^{+0.015} _{-0.003}$ & $ ^{+0.024} _{-0.001} $ & 2.44 & 0.772 \\
\hline
60  & 0.00150 & $0.303$ & $ \pm 0.046$ & $ ^{+0.089} _{-0.033}$ & $ ^{+0.012} _{-0.016} $ & 1.84 & 1.384 \\
    & 0.00300 & $0.259$ & $ \pm 0.026$ & $ ^{+0.055} _{-0.020}$ & $ ^{+0.009} _{-0.008} $ & 1.54 & 1.107 \\
    & 0.01200 & $0.109$ & $ \pm 0.011$ & $ ^{+0.009} _{-0.005}$ & $ ^{+0.015} _{-0.002} $ & 2.24 & 0.710 \\
\hline
130 & 0.00300 & $0.214$ & $ \pm 0.054$ & $ ^{+0.071} _{-0.024}$ & $ ^{+0.009} _{-0.018} $ & 1.60 & 1.290 \\
    & 0.00600 & $0.287$ & $ \pm 0.041$ & $ ^{+0.012} _{-0.045}$ & $ ^{+0.012} _{-0.010} $ & 1.51 & 1.005 \\
    & 0.03000 & $0.065$ & $ \pm 0.012$ & $ ^{+0.030} _{-0.010}$ & $ ^{+0.008} _{-0.002} $ & 2.51 & 0.575 \\
\hline
500 & 0.01200 & $0.338$ & $ \pm 0.065$ & $ ^{+0.029} _{-0.092}$ & $ ^{+0.021} _{-0.024} $ & 1.57 & 0.905 \\
    & 0.03000 & $0.180$ & $ \pm 0.041$ & $ ^{+0.023} _{-0.026}$ & $ ^{+0.012} _{-0.005} $ & 2.42 & 0.624 \\
\hline

\end{tabular}
\caption{\textit{
The extracted values of $F_2^{c\bar{c}}$ at each $Q^2$ and $x$ value. The statistical, systematic 
and theoretical uncertainties are shown separately. The values of the extrapolation factor used 
to correct the full \ptds\ and \etads\ phase space are also shown. The value of the proton structure 
function, $F_2$, from the ZEUS NLO QCD fit used to extract the ratio $F_2^{c\bar{c}}/F_2$, is also 
given.}}
\label{tab:f2c_bins}
\end{center}

\end{table}



\newpage

\newpage

\newpage

\newpage

\newpage

\newpage

\newpage

%
%
\end{document}